# Liquid-like correlations in single crystalline $Y_2Mo_2O_7$: an unconventional spin glass

**H.J. Silverstein[1*], K. Fritsch[2], F. Flicker[3, 4, 5], A.M. Hallas[1], J.S. Gardner[6, 7], Y. Qiu[7, 8], G. Ehlers[9], A.T. Savici[10], Z. Yamani[11], K.A. Ross[2], B.D. Gaulin[2, 12, 13], M.J.P. Gingras[3, 12], J.A.M. Paddison[14, 15], K. Foyevtsova[16], R. Valentí[16], F. Hawthorne[17], C.R. Wiebe[1, 18] and H.D. Zhou[19, 20]**

*****Correspondence to H. J. Silverstein at umsilve3@myumanitoba.ca

*1 – Department of Chemistry, University of Manitoba, Winnipeg, MB, R3T 2N2, Canada*

*2 – Department of Physics and Astronomy, McMaster University, Hamilton, ON, L8S 4M1, Canada*

*3 – Department of Physics and Astronomy, University of Waterloo, Waterloo, ON, N2L 3G1, Canada*

*4 – Perimeter Institute for Theoretical Physics, 31 Caroline Street North, Waterloo, ON, N2L 2Y5, Canada*

*5 – School of Physics, HH Wills Physics Laboratory, University of Bristol, Bristol, BS8 1TL, UK*

*6 – Indiana University, 2401 Milo B. Sampson Lane, Bloomington, Indiana, 47408, USA*

*7 – NIST Center for Neutron Research, National Institute of Standards and Technology, Gaithersburg, MD, 20899-6102, USA*

*8 – Department of Materials Science and Engineering, University of Maryland, College Park, MD, 20742, USA*

*9 – Quantum Condensed Matter Division, Neutron Sciences Directorate, Oak Ridge National Laboratory, Oak Ridge, TN, 37831, USA*

*10 – Neutron Data Analysis and Visualization Division, Neutron Sciences Directorate, Oak Ridge National Laboratory, Oak Ridge, TN, 37831, USA*

*11 – Canadian Neutron Beam Centre, National Research Council, Chalk River, ON, K0J 1P0, Canada*

*12 – Canadian Institute for Advanced Research, 180 Dundas St. W., Toronto, ON, M5G 1Z8, Canada*

*13 – Brockhouse Institute for Materials Research, McMaster University, Hamilton, ON, L8S 4M1, Canada*

*14 – Department of Chemistry, University of Oxford, Inorganic Chemistry Laboratory, South Parks Road, Oxford OX1 3QR, UK*

*15 – ISIS Facility, Rutherford Appleton Laboratory, Chilton, Didcot, Oxfordshire OX11 0QX, United Kingdom*

*16 – Institut für Theoretische Physik, Goethe-Universität Frankfurt, Frankfurt am Main, 60438, Germany*

*17 – Department of Geological Sciences, University of Manitoba, Winnipeg, MB, R3T 2N2, Canada*

*18 – Department of Chemistry, University of Winnipeg, 515 Portage Ave, Winnipeg, MB, R3B 2E9, Canada*

*19 – Department of Physics and Astronomy, University of Tennessee, Knoxville, TN, 37996-1200, USA*

*20 – National High Magnetic Field Laboratory, Florida State University, Tallahassee, FL, 32306-4005, USA*

**Abstract**

The spin glass behavior of $Y_2Mo_2O_7$ has puzzled physicists for nearly three decades. Free of bulk disorder within the resolution of powder diffraction methods, it is thought that this material is a rare realization of a spin glass resulting from weak disorder such as bond disorder or local lattice distortions. Here, we report on the single crystal growth of $Y_2Mo_2O_7$. Using neutron scattering, we present unique isotropic magnetic diffuse scattering arising beneath the spin glass transition despite having a well-ordered structure at the bulk level. Despite our attempts to model the diffuse scattering using a computationally exhaustive search of a class of simple spin Hamiltonians, we were unable to replicate the experimentally observed energy-integrated (diffuse) neutron scattering. A $T^2$-temperature dependence in the heat capacity and density functional theory calculations hint at significant frozen degeneracy in both the spin and orbital degrees of freedom resulting from spin-orbital coupling (Kugel-Khomskii type) and random fluctuations in the Mo environment at the local level.

## I. INTRODUCTION

The spin glass phase is just one of the many magnetic states that can arise from competing interactions resulting from spins residing on a geometrically frustrated lattice. Spin glassiness is characterized by disordered moments frozen in time along different orientations [1]:

$$1/N \sum_i \langle S_i \rangle_{t'} e^{(i\mathbf{Q}\cdot\mathbf{R}_i)} = 0 \qquad (N \to \infty),\ \langle S_i \rangle_{t'} \neq 0$$

where $\langle S_i \rangle_{t'}$ denotes a macroscopic time average over a single spin. Since the 1970s, it has been shown that the spin glass state is quite common among materials with random disorder, occurring in dilute doped metals like AuFe [2], non-metallic solid solutions $Eu_xSr_{1-x}S$ [3], stoichiometric crystalline materials such as $Fe_2TiO_3$ (where site mixing occurs between magnetic $Fe^{3+}$ and nonmagnetic $Ti^{4+}$) [4], and amorphous materials [5]. Although describing these systems theoretically is not trivial, some models, especially those that build upon the Edwards-Anderson model [6], have proven quite successful to flesh out the essential physics governing spin glass behavior [1]. Difficulties arise because the correct definition of an order parameter for a three-dimensional spin glass and the existence of clear universality classes remain a matter of debate. The degeneracy, at least approximate, of the low-energy states in these systems arises from a “many-well” free energy landscape. For such a non-trivial many-well landscape to exist and drive a genuine thermodynamic spin glass state, as opposed to order-by-disorder [7, 8], the Hamiltonian must contain two essential ingredients: competition between the terms in the Hamiltonian such that all terms cannot be minimized simultaneously (known as frustration) and randomness, usually caused through chemical disorder. The typical spin glass can be recognized from some key experimental signatures including a zero-field cooled/field-cooled split in the magnetic susceptibility at the freezing temperature $T_f$, a frequency dependence in the AC susceptibility, a broad peak in the heat capacity occurring near $T_f$, a large relaxation of the spin dynamics as $T_f$ is approached, and no onset of long-range magnetic order as probed by neutron scattering. Yet perhaps the clearest indicator of thermodynamic spin glass freezing is the divergence of the nonlinear magnetic susceptibility [1]. Over the past three decades, all of these experimental signatures have been observed in the magnetic pyrochlore $Y_2Mo_2O_7$.

The magnetic pyrochlores have been an immense source of new and exotic physics at low temperatures [9] with some notable examples including the spin ice state in $A_2X_2O_7$ (A = $Ho^{3+}$, $Dy^{3+}$; X = $Ti^{4+}$, $Ge^{4+}$) [10-13], potential spin liquid states in $Tb_2Ti_2O_7$ [14, 15] and $Er_2Sn_2O_7$ [16], order by disorder mechanism in $Er_2Ti_2O_7$ [17-19], dynamic spin ice states in $Pr_2Zr_2O_7$ [20] and $Pr_2Sn_2O_7$ [21], and even quantum spin ice state $Yb_2Ti_2O_7$ [22-23] although controversy still surrounds the ground state of the latter. The pyrochlore structure (Z=8, space group:$Fd\bar{3}m$) typically contains an 8-fold coordinated trivalent cation surrounded by 6-fold coordinated tetravalent cations (Figure 1a), although defects are known to occur [9]. For the purposes of this study, it is quite convenient to envision the pyrochlore as two ordered interpenetrating networks of corner-linked tetrahedra with cationic vertices (Figure 1b). This highly frustrated sublattice topology is the culprit behind the richness of the magnetic ground states (Figure 1b, c).

Some of the earlier reported syntheses of $Y_2Mo_2O_7$ (S=1) pyrochlore stem back to 1980 [24] where the electrical properties and specific heat [25] were examined in conjunction with other pyrochlores. It was first shown by Reimers *et al.* and Greedan *et al.* [26, 27] that $Y_2Mo_2O_7$ displays all the characteristics of typical spin glass behavior below the freezing temperature of $T_f$=22.5 K [9, 27-30]. Nearly two decades later, $Y_2Mo_2O_7$ was shown to display critical behavior characteristic of randomly disordered spin glasses derived from the scaling of the non-linear susceptibility [31], despite missing one of the key ingredients of the spin glass Hamiltonian: chemical disorder. Here, the term “chemical disorder” includes site-mixing, oxygen vacancies, and nonstoichiometry - all of which are negligible within the resolution of X-ray and neutron powder diffraction. The properties of $Y_2Mo_2O_7$ have been studied in detail using just about every well-established experimental technique available including bulk probes like neutron spectroscopy [28, 32], DC and AC magnetic susceptibility [30, 33] and resistivity [24, 34], as well as with local probes like NMR [35, 36] and μSR [29]. $Y_2Mo_2O_7$ is a large band-gap semiconductor ($E_{gap}$ = 0.013 eV) with strong $Mo^{4+}$-$Mo^{4+}$ spin interactions (the Curie-Weiss temperature, $\theta \approx -200$ K). Neutron diffraction revealed the absence of any magnetic Bragg peaks, although magnetic diffuse scattering develops beneath $T_f$ centered at $\mathbf{Q}$ = 0.44 $Å^{-1}$. This feature has been interpreted as spin correlations occurring over four sub-lattice structures along the <110> directions [28]. It was also shown that the inelastic neutron scattering spectral weight completely vanishes within the instrumental resolution as the temperature is lowered towards $T_f$. Evidence from μSR spectroscopy, which probes fluctuation rates in the range of $10^4$-$10^{11}$ $s^{-1}$ [29], finds a complex internal magnetic field distribution below $T_f$ that prevents coherent muon spin precession. This is accompanied by a simultaneous power-law decrease in the spin relaxation rate. Evidence for residual spin dynamics at $T/T_f < 0.05$ was also found for this system and is generally atypical of spin glasses [29]. However, the number of studies on $Y_2Mo_2O_7$ at these temperatures is quite limited and these residual spin dynamics will not be considered further here. NMR experiments revealed a dramatic increase in the number of discrete $^{89}Y$ sites as the temperature was cooled to 77 K [35] (three times higher than $T_f$). Here, the authors of [35] reason that a frustration-driven lattice distortion is responsible for the increase in the number of sites, although they were not able to explain the discreteness of the sites.

Whereas bulk probes indicate negligible disorder within the resolution of the experiment, the presence of (and effects due to) disorder are more obvious with the local probes. Local probes like extended X-ray absorption fine structure spectroscopy (EXAFS) [37], neutron pair distribution function (nPDF) analysis [38], and the previously discussed μSR and NMR studies paint a very different picture of $Y_2Mo_2O_7$. For example, EXAFS indicate the presence of small amounts of bond randomness to varying degrees in powder samples of $Y_2Mo_2O_7$, particularly in the Mo-Mo bond distance. In contrast, nPDF analysis highlights the important role of large anisotropic variations in the Y-O1 bond lengths (it should be noted that fits to the data worsened as the temperature was cooled, contrary to what would be expected as the thermal contribution to the scatter decreases). Although there is a consensus that the level of bond disorder in this material is rather low, there is no agreement on what that level of disorder, or even type of disorder, might be.

It is worth quickly taking note of some other magnetic molybdate systems. The physics present in molybdate pyrochlores varies widely, especially across the rare-earth containing molybdate pyrochlores, due to the proximity of a Mott transition. Unlike $Y_2Mo_2O_7$, $A_2Mo_2O_7$ (A=Nd, Sm, Gd) are all metallic ferromagnets where Mo moments order at $T_c$=97, 93, and 83 K respectively [39-41]. It is known that ferromagnetism and metallic character in these pyrochlores are tightly linked due to spin-orbit coupling and resultant splitting of the Mo crystal fields, despite subtle structural changes as one moves across the series [9, 42]. Nd and Sm molybdate pyrochlores also exhibit the anomalous Hall effect, although a clear explanation for the mechanism behind this behavior remains to be seen [43, 44]. $Tb_2Mo_2O_7$ is a spin glass with two magnetic ions which lies close to the metal-insulator phase boundary [29]. $Lu_2Mo_2O_7$ is a relatively new pyrochlore that appears to have similar properties to $Y_2Mo_2O_7$, although more study is necessary [45]. Pyrochlore antimonides are also intriguing, behaving very similarly to $Y_2Mo_2O_7$, but the complex synthesis has made these systems a somewhat less attractive avenue of study [46, 47]. Other Y-Mo containing systems include the double perovskite $Ba_2YMoO_6$ [48-50]. The true ground state of the $Ba_2YMoO_6$ system is still a matter of debate, with some reports pointing towards a frozen valence bond glass (a frozen disordered pattern of spin singlets) [48, 49] while others suggest a quantum spin-liquid ground state with strong spin-orbit coupling [50].

There are a myriad of theoretical studies on molybdate systems and a thorough review of all of them is beyond the scope of this text [9]. Instead, we restrict ourselves to theoretical investigations regarding pyrochlore Heisenberg antiferromagnet spin glasses with weak disorder. Even with our restriction, theoretical examinations concerning the possibility of weak disorder-induced spin glass states in pyrochlores are still quite numerous. $Y_2Mo_2O_7$ is thought to be well-represented by a Heisenberg antiferromagnet on the pyrochlore lattice, and while a defined spin glass transition occurs in $Y_2Mo_2O_7$, no such transition is expected for this model at any temperature. Bellier-Castella *et al.* [51] found that bond-disorder can lift the degeneracy expected for this system and induce a short-ranged ordered collinear spin structure, similar to what was proposed from neutron scattering experiments [28]. However, the energy scale of this process is orders of magnitude lower than that determined from experiment. Andreanov *et al.* [52] and Saunders and Chalker [53] also explored the possibility of exchange randomness as a means of lifting the degeneracy in the perfectly-ordered limit. Here, they determined that variations in exchange create long-range effective couplings that induce spin freezing at a temperature set by the strength of the disorder, although whether or not this mechanism is true for $Y_2Mo_2O_7$ is still ambiguous. Tam *et al.* [54] suggest that the spin glass state is not caused by weak-disorder, but rather very strong effective disorder possibly due to perturbations beyond nearest-neighbor exchange. Furthermore, Shinaoka *et al.* [55, 56] note that if $T_f$ is necessarily set by the strength of disorder, then partial substitution of $Y^{3+}$ for $La^{3+}$ should result in an increase in $T_f$, which is not what is observed experimentally despite a significant change in the θ [57]. Rather, Shinaoka *et al.* suggest local lattice distortions from spin-lattice coupling as a mechanism for spin freezing [55, 56].

While a clear solution to the $Y_2Mo_2O_7$ spin glass problem may yet take years to surface, one issue that has stalled progress is that all of the aforementioned measurements were made on powder samples instead of single crystals. The lack of three-dimensional **Q**-space information in powders is the largest hindrance to understanding the magnetism of *any* system. In order to address this problem and move the field forward, we report to our knowledge the first single crystal of $Y_2Mo_2O_7$ using the optical floating zone technique. The growth of such a crystal is not trivial: in general, the molybdates are very difficult to crystallize due to the rapid oxidation of $Mo^{4+}$ to non-magnetic $Mo^{6+}$ at high temperatures. For the case of $Y_2Mo_2O_7$, an additional inhibitor to crystal growth includes an inherent electronic instability due to the proximity of the metal-insulator transition (this has been observed experimentally in this system by doping small amounts of Cd on the Y site [58]). We first address concerns of crystal quality with the use of X-ray diffraction, electron microprobe analysis and magnetic susceptibility, the latter of which is highly susceptible to oxygen nonstoichiometry. Results from elastic and inelastic neutron scattering experiments are presented along with synchrotron X-ray scattering experiments. Our attempts at modeling the diffuse magnetic scattering are then discussed. Extensive heat capacity experiments are next presented, which shed new light on the $Y_2Mo_2O_7$ problem. Finally, we will show that density-functional theory calculations support our claim that the degeneracy in $Y_2Mo_2O_7$ is found not only in the spin system, but in the orbital system as well as a result of strong spin-orbital-lattice coupling. We note here that spin-orbital coupling does not refer to the relativistic atomic spin-orbit (S·L) coupling, rather the multidimensional spin-orbital coupling of the Kugel-Khomskii type derived from the Hubbard model [59].

## II. EXPERIMENTAL

### A. Sample preparation and crystal growth

$Y_2O_3$ (99.99%) and $MoO_2$ (99.9%) powders were ground in stoichiometric amounts, pelleted, and sintered at 1425 K for 48 hr under flowing $N_{2(g)}$ with intermittent grindings. A final reduction step using $H_{2(g)}$ and loose powder sample was done in order to obtain phase pure powder $Y_2Mo_2O_{7,}$ which was verified with an initial X-ray powder diffraction measurement. Single crystal growth was done at the National High Magnetic Field Laboratory (NHMFL, Tallahassee, FL). Powder $Y_2Mo_2O_7$ was pressed into 6-mm diameter 60-mm rods under 400 atm hydrostatic pressure and calcined in Ar at 1400 K for 12 hours. In order to compensate for the loss of $MoO_3$ during the growth, 15% excess $MoO_2$ was added to the rods. The crystal growth was carried out in $Ar_{(g)}$ in an infrared heated image furnace equipped with two halogen lamps and double ellipsoidal mirrors with feed and seed rods rotating in opposite directions at 25 rpm during the crystal growth at a rate of 30 mm/hr. Controlling the hot zone is extremely difficult: once the proper conditions were found, over 15 growths were attempted. All crystals were annealed in temperatures below 700°C in a $CO/CO_2$ buffer gas to compensate for oxygen nonstoichiometry, until the glassy transition temperature remained stable (this usually involved heating the samples overnight). The final crystal used for all neutron scattering measurements was approximately 3 cm in length and 0.3 cm in diameter. Powder $Y_2Ti_2O_7$ was synthesized by methods previously reported in the literature [60].

### B. Crystal characterization

In the case of $Y_2Mo_2O_7$, X-ray Rietveld refinement is not the best tool for characterization of essential structural issues (in this case, we are concerned mainly with Y/Mo site mixing and O nonstoichiometry). Assuming powder $Y_2Mo_2O_7$ is a well-ordered, stoichiometric reference standard (a good assumption considering that the wealth of diffraction, magnetization, and heat capacity measurements performed across many samples is largely consistent), the DC magnetic susceptibility, particularly θ and $T_f$, are a much better indicator of sample quality, as both are sensitive to nonstoichiometry, site-mixing, crystallite size, and unit cell size [61-66]. The best crystal was selected based on the consistency of the shape of the DC magnetic susceptibility curve, $T_f$, and θ with previous powder samples. Additionally, we performed single-crystal X-ray diffraction, Rietveld refinement, and back-scattered electron imaging (BSE) using wavelength-dispersive electron microprobe analysis, all of which were performed at the University of Manitoba (Winnipeg, MB) using a polished 0.03 g cross-section of the crystal. During our elemental analysis, a sputtering gun was used to probe the interior of the crystal, which left small Mo metal inclusions on its surface (as seen in Figure 2). The magnetic susceptibility was measured as a function of temperature using a superconducting quantum interference device (SQUID) with applied fields of 0.1 T up to 5 T applied along the [111] direction.

**C. Scattering experiments**

Time-of-flight neutron spectroscopy was performed on the Disc Chopper Spectrometer (DCS) [67] at the NIST Center for Neutron Research (NCNR, Gaithersburg, MD) using wavelength λ =4.8 Å and at the Cold Neutron Chopper Spectrometer (CNCS) [68] at the Spallation Neutron Source (SNS, Oak Ridge, TN) using neutrons of incident energy $E_i$=3, 20 meV. Measurements from the triple axis instrument were taken on C5 at the Canadian Neutron Beam Centre (CNBC, Chalk River, ON) using vertically focusing PG002 monochromator and flat PG002 analyzer crystals with fixed final energy $E_f$=3.52 THz, a single filter, and a [none, 0.8°, 0.85°, 2.4°] collimation setting. Mesh scans were created using a series of line scans along [HH0] over [00L]. All neutron measurements were made over the temperature interval [1.5 K, 300 K]. X-ray measurements were performed on a 2-circle laboratory source equipped with 14.4 keV X-rays and at the 4-ID-D beamline at the Advanced Photon Source (Argonne, IL) in transmission geometry on a ~200 micron thick and polished sample. 16.9 keV X-rays were used to avoid the absorption edges of Y and Mo. These measurements were made over the temperature interval [10 K, 300 K]. All crystal alignments were done at McMaster University (Hamilton, ON). The Horace (http://horace.isis.rl.ac.uk/Main_Page) and DAVE packages were used for data analysis [69].

**D. S(Q) modeling**

The large $N$ method in the present context entails enlargement of the symmetry group of the Heisenberg spins from $O(3)$ to $O(N)$. In the limit $N \to \infty$, the corresponding Hamiltonian is exactly solvable [70]. An expansion in $1/N$ about $1/N = 0$ could then be systematically carried out [71, 72]. The resulting diffuse scattering factor $S(\mathbf{Q}) = \sum_{i,j=1}^{4} [(\lambda \mathbb{I}_4 - \beta \sum_n J_n A^{(n)}(\mathbf{Q}))^{-1}]_{ij}$ with $\mathbb{I}_4$ the 4 x 4 identity matrix, $\beta$ the inverse temperature, $\lambda$ a Lagrange multiplier constraining the average spin length to be $\langle S \rangle = 1$, and $A^{(n)}(\mathbf{Q})$ the Fourier transform of the $n^{\text{th}}$ nearest neighbor structure [70]. To first order in $J$ we have that $4(J_1 + 2J_2 + J_{3a} + J_{3b} + 2J_4) = -200$ K (noting that there are two inequivalent types of third nearest neighbors in a pyrochlore lattice, which we assumed had equal couplings [73]). The search of this parameter space was carried out numerically using a FORTRAN code written by the authors.

Additionally we derived an analytic expression for the scattering along [00L] then searched numerically for maxima lying at **Q** = 0.44 Å$^{-1}$. An alternative approach to calculating the diffuse scattering factor is to maintain $O(3)$ spins and employ a mean-field approximation, as detailed in [74]. This approach yields diffuse scattering factors in qualitative agreement with the spherical approximation.

**E. Heat capacity measurements**

Heat capacity measurements were made using a Physical Property Measurement System in both 0 field and 9 T field applied along the [111] direction. These measurements were made at the NHMFL and at the University of Winnipeg. For these measurements, an approximately 16 mg subsection of the single crystal $Y_2Mo_2O_7$ was measured for reproducibility, and was then crushed with a mortar and pestle for 30 minutes. The powder was pelleted before the heat capacity was measured. The final mass of each pellet is about 7-8 mg, but is only known to within 1 mg. Finally, the pellet was annealed at 300°C for 24 hours in $O_{2(g)}$ and the heat capacity was once again measured. This was done to test for $O_2$ surface effects. Since the original preparation of the material is typically done in reducing atmospheres or under vacuum, it was felt that 300°C was a fair choice of temperature, acting as a compromise between $O_2$ reactivity with the surface and $MoO_3$ volatility loss. The heat capacity of $Y_2Ti_2O_7$ was taken from the values reported in the literature [60] and confirmed with the heat capacity of our own samples of $Y_2Ti_2O_7$.

**F. Density function theory calculations**

The density functional theory electronic structure calculations were performed using the projector augmented wave method [75] as implemented in the Vienna *ab initio* simulation package (VASP) [76-79] and the linearized augmented plane wave method as implemented in the full-potential *ab initio* code Wien2k [80, 81]. The exchange-correlation functional is described within the spin-polarized generalized gradient approximation (GGA) [82]. In order to reproduce the Mott insulating state in $Y_2Mo_2O_7$, we add an orbital dependent term to the GGA functional that mimics the on-site Coulomb repulsion $U$ between Mo 4*d* electrons, following the self-interaction correction scheme by Liechtenstein *et al.* [83] in Wien2k and the Dudarev *et al.* scheme [84] in VASP. The Hund's exchange coupling and the on-site Coulomb repulsion are set, respectively, to 0.5 eV and 4 eV. In order to be able to capture possible orderings of the Mo 4*d* orbitals, we switch off symmetrisation in VASP and consider a reduced $Y_2Mo_2O_7$ unit cell with artificially lowered symmetry *P*-1 in WIEN2k. For the structural input we use powder neutron diffraction data reported in [26]. Results of the structural relaxation were cross-checked by relaxing internal parameters of a cubic unit cell containing 88 symmetry unrestricted atoms for the AFM-OO2 state.

## III. RESULTS AND DISCUSSION

**A. Initial characterization and magnetic susceptibility**

The growth of any single crystal pyrochlore molybdate is a tremendous endeavor. For example, single crystals of $Gd_2Mo_2O_7$, $Sm_2Mo_2O_7$ and $Nd_2Mo_2O_7$ have been grown [9, 85-87]. In addition to the volatility of $MoO_3$ at crystal growth temperatures, one has to be cautious of oxygen nonstoichiometry.

In the case of $Gd_2Mo_2O_7$, a molybdate pyrochlore particularly close to the metal-insulator phase boundary, oxygen nonstoichiometry causes some $Mo^{4+}$ to reduce to larger $Mo^{3+}$, resulting in greater Mo-O distances and a larger lattice constant which in turn causes the material to cross into the insulating regime from the metallic one [9]. This can be fixed simply by annealing $Gd_2Mo_2O_7$ in a CO/$CO_2$ buffer gas – a step that we have taken here. Such dramatic effects on the properties are not observed for $Sm_2Mo_2O_7$ and $Nd_2Mo_2O_7$ crystals as they are farther from the phase transition, although the Curie and Weiss temperatures obtained from a Curie-Weiss law fit to the magnetic susceptibility deviated from stoichiometric powder values [9, 86-88].

Figure 2 displays a back-scattered electron image taken using wavelength-dispersive electron microprobe analysis. The white areas are Mo metal inclusions on the surface of the cross section resulting from a sputtering process used to probe the interior. This particular piece of crystal used in the characterization analysis was not used in any other measurement – therefore, a Mo metal impurity does not appear in any synchrotron, neutron, magnetic susceptibility, or heat capacity measurement. The grey area is stoichiometric $Y_2Mo_2O_7$. Figure 3 shows a composite image of single crystal X-ray diffraction scans on a single grain 0.1 mm in size, while Figure 4 displays single exposures from a charge-coupled device. Mo metal surface inclusions are again visible as arcs in this figure. The orthogonal array of diffraction spots is from the dominant $Y_2Mo_2O_7$ phase. Our single crystal – in the absolute strictest sense – is actually a *small* collection of F-centered crystallites in different orientations (a=10.28 Å), all of which can be indexed to previously published $Y_2Mo_2O_7$ refinement data (ICSD 202522, Fd-3m, a=10.23 Å, R=2.4%). It is believed that the small difference in the lattice parameter is due to the composite nature of our crystal rather than compositional or structural defects at the unit-cell level. At the bulk level, there is one major grain and two minor grains in our sample as detected with neutron scattering (not shown). A crystal growth of this quality is typical of Image Furnace growths: in general, it is extremely difficult to obtain a true single crystal of this size free of internal grain boundaries using this method. However, for the experimental purposes described in this text largely focused on the bulk properties of the material, the composite can be approximated as one single crystal.

On the other hand, magnetic susceptibility is an excellent indicator of molybdate pyrochlore crystal quality. Figure 5a shows that a field-cooled/zero field-cooled split in the susceptibility, characteristic of other spin-glasses, is observed in our single crystal at the transition $T_f$ = 22.5 K. This is consistent with all other reported measurements on powder samples of $Y_2Mo_2O_7$ [9,26-28, 31]. The Curie-Weiss law was used to fit the inverse DC-susceptibility where an effective moment of $\mu_{eff}$=2.1±0.1 $\mu_B$ was found within the 50-300 K temperature range. This is lower than the theoretical $Mo^{4+}$ moment ($\mu_{eff}$=2.83 $\mu_B$) but also consistent with powder samples. The Curie-Weiss temperature was found to be θ =-200 K (with no change in the effective moment within error) when the region of fit was extended to 600 K (not shown), the highest temperature that would be allowed without sample degradation, which is consistent with previous powder samples [28]. However, the Curie-Weiss temperature becomes more ferromagnetic as the region of fit was reduced to lower temperatures eventually reaching a minimum of θ =-41 K. This is also consistent with values found in the literature for polycrystalline samples [9] indicating that our sample is consistent with previous powder samples *in both* fitting regimes. Again, we stress that the consistency of the DC susceptibility is a direct and non-destructive indication of the consistency in the quality between Mo pyrochlore samples, as is also demonstrated in a very recent study on isostructural $Lu_2Mo_2O_7$ [45]. Recently, there has been a theoretical treatment of a crossover region between two Curie-Weiss regimes as it relates to topological sector fluctuations in $Ho_2Ti_2O_7$ spin ice [89]. Although it is conceivable that this might potentially relate to the present system, it must be remembered that the definition of such a topological sector in spin ice is aided by strict adherence to the so-called 'ice rules' – a feature absent here. There is no evidence from magnetic susceptibility to support $Y_2Mo_2O_7$ entering into a different regime from the Heisenberg antiferromagnet state, rather the spins merely freeze out. A prominent cusp in the susceptibility at the freezing temperature is observed in Figure 5b, which broadens and moves to higher temperatures with a field applied along [111]. This is typical of spin glasses in general [1].

**B. Neutron and synchrotron X-ray scattering**

Neutron scattering experiments to measure the diffuse scattering at low temperatures were completed at three neutron sources (the DCS at NIST, C5 at Chalk River, and CNCS at ORNL). All three sources confirmed the presence of significant diffuse scattering at low temperatures – in particular, an elastic feature centered at $\mathbf{Q}$ = 0.44 $Å^{-1}$ (Figure 6a with cuts along various directions are shown in Figure 7) - without the appearance of magnetic Bragg peaks whilst using a high temperature data set for subtraction. The diffuse scattering is likely the same feature reported by Gardner *et al.* in powder samples [28]. Remarkably, this "ring" feature is completely isotropic in $\mathbf{Q}$, as is evident from the cuts along various planes of symmetry in Figure 7, reminiscent of liquid-like scattering. The width of this ring feature is beyond the Q-resolution limit of the DCS and yields a correlation length of 5.3±0.5 Å estimated from the inverse half-width at half-maximum from a Gaussian fit that agrees with earlier measurements on powders [28]. For net antiferromagnetic interactions, we would not expect this pattern of diffuse scattering. For example, Zinkin *et al.* calculated a distinct $\mathbf{Q}$-dependence for 3D Heisenberg spins on the pyrochlore lattice marked by the absence of $\mathbf{Q}$ = 0 scattering, which does not resemble our data at all [90]. Nor does our scattering resemble that predicted by Moessner and Chalker [91], Conlon and Chalker [72], or Henley [71] for various states in the pyrochlore Heisenberg antiferromagnet. Indeed, our "rings" appear around the origin and other ferromagnetic points such as {222} in the next Brillouin zone. The ring and (222) magnetic scattering, observed as peaks along the [HHH] direction in Figure 6b, both appear 0.44 $Å^{-1}$ away from the peak center and share a Lorentzian-like tail radiating outward. Ring-like diffuse scattering in single crystalline materials has been observed in fast ion conductors such as α-AgI [92], where liquid-like correlations are expected due to mobile $Ag^+$ ions trapped in an $I^-$ ion network. There have also been reports of magnetic rings in MnSi [93, 94] due to skyrmions using small angle neutron scattering, as well as due to magnetic short range order in $Nd_3Ga_5SiO_{14}$ [95] (which was caused by liquid He leakage from the cryostat and was redacted by the authors [96]). Rings have also been reported as part of a larger pattern of excitations in $ZnCr_2O_4$ [97] and $MgCr_2O_4$ spinels [98], but to our knowledge, elastic *magnetic* rings of the sort observed in our study *have never been reported*. While some similarities to skyrmionic systems can be drawn, particularly the incommensurate nature of the magnetic rings, a key difference is that the momentum vector of the ring observed here corresponds to 4 sub-lattice correlations *within the unit cell*. The typical size of a skyrmion is, at minimum, an order of magnitude larger than what is calculated for $Y_2Mo_2O_7$. So far to our knowledge, there is no reported evidence or predictions for such a state in $Y_2Mo_2O_7$, or any pyrochlore system, and this idea will not be considered further here. No predictions have been made in the literature of an isotropic magnetic ring of scattering of any sort in a pyrochlore, although such a "ring liquid" has been predicted for the honeycomb lattice using a two-neighbor antiferromagnetic exchange [99]. Figure 6c displays a broad, low-momentum excitation in the inelastic channel at high temperatures that vanishes as $T_f$ is approached, which is consistent with results from Gardner *et al.*, 1999 [28]. Unfortunately, more work is required to characterize these excitations in any quantifiable context. What is known is that, like the ring, the scattering of these excitations is also isotropic. Unlike the ring, however, there is no indication near **Q** = 0 of the disappearance of scatter while the dispersion of these excitations exist up until about **Q** = 0.25 $Å^{-1}$ instead of **Q** = 0.44 $Å^{-1}$ (please refer to the figures in the supplementary information). Figure 6d compares the raw scattering at different temperatures (two small peaks observed at (0.6, 0.6, 0.6) and (1.4, 1.4, 1.4) are instrumental artifacts). Here, extra magnetic scattering is clearly visible surrounding the (222) peak below the freezing temperature.

Using the CNCS, we investigated the elastic region with neutrons of incident energy $E_i$ = 20 meV at both 300 K and 1.5 K. This allowed us access to regions farther out in **Q** in the HHL plane like (008) and (440). In particular, half-butterfly regions of diffuse scattering appear around these two Bragg peaks, reminiscent of Huang scattering. Data collection was obtained over similar time intervals and all intensities were normalized to the source flux. A clear decrease in the overall intensity is observed in Figure 8 as the temperature is raised from 1.5 K to 300 K. The Huang scattering patterns and the intensity difference were both verified using the C5 triple axis spectrometer at Chalk River (inset, high temperature not shown). However, we are unable to comment on the temperature dependence trend of the Huang scattering due to a lack of data at intermediate temperatures. The persistence of this diffuse scattering to temperatures much higher than θ motivated us to investigate the scattering at the Advanced Photon Source on the 4-ID-D beamline. A representative pattern is shown in Figure 8 for the (660) peak, which was not accessible on the CNCS using $E_i$=20 meV. We note however that Huang scattering was also observed for the (440), (880), (008), and (666) peaks, all of which were many times more intense than that shown for the (660) peak (part of the (880) scattering can be observed in Figure 9 on the right). The Huang scattering was not observable with conventional laboratory sources. No obvious temperature dependence for the scattering was observed, unlike what was seen with neutron scattering.

Huang scattering is observed in many chemically disordered systems such as $Fe_{1+x}Te$, $YBa_2Cu_3O_{6.92}$ and $La_{1.2}Sr_{1.8}Mn_2O_7$ [100-102]. In $La_{1.2}Sr_{1.8}Mn_2O_7$, the Huang diffuse scattering is due to Jahn-Teller distortions, which also play a role in the colossal magnetoresistance seen in this material. Many factors can cause Huang scattering, and the origin in $Y_2Mo_2O_7$ is still speculative at best. For example, chemical disorder due to oxygen deficiencies may cause Huang scattering by introducing variances in the local charge, but this is unlikely in $Y_2Mo_2O_7$; the magnetic susceptibility, glassy transition temperature and diffuse scattering, all of which are highly influenced by non-stoichiometry and site-mixing, are consistent with well-ordered powder samples with respect to oxygen. For now, the origin of the butterfly patterns in $Y_2Mo_2O_7$ remains unclear although we offer some speculation in sections III, D and E.

**C. S(Q) modeling**

The simplest approach to modeling the Hamiltonian responsible for the magnetic interactions is with nearest neighbor $Mo^{4+}$-$Mo^{4+}$ exchange. Many attempts to model the magnetic diffuse scattering were made using a variety of isotropic spin Hamiltonians. Using a large $N$ method [70] applied to Heisenberg spins, we carried out an exhaustive numerical and analytical test of possible spin-spin couplings out to the fourth nearest-neighbor, which is well beyond the correlation length calculated by Gardner *et al.* [28] and confirmed here (not shown).

The spins of the $Mo^{4+}$ ions in $Y_2Mo_2O_7$ are of the Heisenberg type, having continuous symmetry group $O(3)$. The idea of a large $N$ expansion is to consider higher dimensional spins with symmetry group $O(N)$. In the limit $N \to \infty$, the corresponding Hamiltonian is exactly solvable [70]. The zeroth order in a $1/N$ expansion has been demonstrated to work well for the case of $N = 3$ [72], and we adopt this approximation here.

The calculated neutron scattering factor is [71, 72]:

$$S(\mathbf{Q}) = \sum_{i,j=1}^{4} \left[ \left( \lambda \mathbb{I}_4 - \beta \sum_n J_n A^{(n)}(\mathbf{Q}) \right)^{-1} \right]_{ij}$$

with $J_n$ the coupling between $n^{th}$ nearest neighbors and $A^{(n)}(\mathbf{Q})$ the Fourier transform of the $n^{th}$ nearest neighbor structure. The indices $i$ and $j$ run over the sites of the pyrochlore's tetrahedral sublattice. The Lagrange multiplier $\lambda$ (solved numerically) enforces the self-consistency condition that the average length of each spin component be one third:

$$\frac{1}{3} = \frac{1}{4N} \sum_{\mathbf{Q} \in BZ} \mathrm{Tr} \left[ \lambda \mathbb{I}_4 - \beta \sum_n J_n A^{(n)}(\mathbf{Q}) \right]^{-1}.$$

Keeping up to 3rd nearest neighbors we searched the parameter space $J_2$, $J_3$ ∈ [-30 K, 10 K] in steps of 2 K for a ring of scattering of radius ≈ 0.44 in the HHL plane. We constrained $J_1$ using the Curie-Weiss temperature, giving $J_1$ ∈ [70 K, -90 K]. Searching the range of temperatures T = 10 K, 50 K, 100 K, 150 K, and 200 K, no ring was found. To give a feel for the evolution of the scattering patterns with $J_n$ a tabulation at T = 150 K is shown in Figure 10.

To avoid the possibility that the ring occupied too small a region of parameter space and was missed by this method, we obtained an analytic expression for $S(\mathbf{Q})$ along the (00L) direction, adding in the 4th nearest neighbors. We searched numerically for $\frac{\partial S}{\partial \mathrm{L}} = 0$ and $\frac{\partial^2 S}{\partial \mathrm{L}^2} < 0$ in L ∈ [0.3, 0.6]. The search covered a large region of parameter space: $J_2$, $J_3$, $J_4$ ∈ [-30 K, 10 K], T ∈ [1 K, 200 K], with the constraint $4(J_1 + 2J_2 + 2J_3 + 2J_4)$ = θ = -200 K.

Due to time and processing limits it was not possible to accurately determine $\lambda$ for each loop of the program. Instead we set the value to a high enough level that it was guaranteed to be larger than the correct value, with the effect being to 'smear out' the scattering pattern in a similar manner to an increased temperature. After many numerical checks, we verified that too high a $\lambda$ would not change the qualitative behavior of $S(\mathrm{L})$; specifically, the zeroes of $\frac{\partial S}{\partial \mathrm{L}}$ are unaffected.

The search routine found some results matching the required conditions. Closer inspection revealed a small maximum around L = 0.5 which was overshadowed by a large peak at L = 2 in each case. To remove these false positives, we adjusted the requirement on the second derivative to give a more pronounced maximum, and further stipulated that there must be no other maxima at higher L. Applying these conditions returned no results in the stated parameter range, suggesting that the model outlined here is insufficient to explain the experimental data.

Ignoring the constraint on θ we can qualitatively reproduce the key features of the experimental data with a number of different choices of exchange constants involving ferromagnetic $J_1$ and net antiferromagnetic further-neighbor interactions. Figure 11a shows the large $N$ scattering pattern for one example of such a model, $\frac{J_2}{J_1} =$ -0.25 (ferromagnetic $J_1$), and Figure 11b shows the mean field theory result for the same parameters. The problem with this is clear: the calculated net interaction is always ferromagnetic whereas the experimental one is antiferromagnetic (as determined by the Curie-Weiss temperature). Net ferromagnetic exchange is impossible to avoid because the ring and diffuse scattering appear only at regions where one would expect ferromagnetic scattering (i.e. surrounding the nuclear Bragg peaks and **Q** = 0). We note, however, that the net ferromagnetic interactions necessitated by the approximate large N and mean-field methods may be a reflection of the upward shift of θ to -41 K as the system enters the strongly correlated spin liquid state at T<<|θ ≈ -200 K|.

**D. Heat capacity**

Heat capacity measurements of single crystal $Y_2Mo_2O_7$, crushed single crystals of $Y_2Mo_2O_7$, crushed single crystals of $Y_2Mo_2O_7$ annealed in $O_{2(g)}$ at 300°C (below the temperature at which $MoO_{3(g)}$ volatilizes), and powder $Y_2Ti_2O_7$ from Johnson *et al.* (2009) [60] are presented in Figure 12a. Figure 12b shows the lattice-subtracted heat capacity of single crystal $Y_2Mo_2O_7$ compared to one of the most recently published measurements on powder samples from Raju *et al.* [30]. Here, $Y_2Ti_2O_7$ was used as a lattice subtraction at low temperatures. This is a suitable approximation as it was shown that the heat capacity of $Y_2Ti_2O_7$, which adopts a cubic pyrochlore structure of similar size and elemental composition to $Y_2Mo_2O_7$, could be attributed to three acoustic modes and 63 nearly dispersionless optical modes per unit cell with no evidence of anomalous lattice dynamics [60]. It is at this point that we see the first discrepancy between measurements done on our single crystal and powder samples: the most striking feature is a $T^2$ dependent low temperature heat capacity in contrast with earlier claims of a linear temperature dependence in powders [30]. Additionally, a broad peak is observed in both powders (Figure 12b) and single crystals (Figure 12c) occurring at 15 K that is nearly independent of magnetic fields, except for a weak feature on the peak shoulder occurring at the glassy transition temperature itself (Figure 12c). Integrating this peak yields an entropy recovery of 14.7% of the theoretical maximum (Figure 12c, inset), suggesting that considerable entropy still remains in this system. Before turning to the cause of the $T^2$ heat capacity, we first address the discrepancy between our heat capacity and previous datasets.

The discrepancy between the temperature dependence of the data sets can be attributed to one of two causes. Firstly, heat capacity measurements on powders have not been published in over 20 years. Instrumentation has since improved in the interim: the spread in the data points is a problem in the powder study and is likely an artifact of a noisy subtraction and differences in instrumentation. A quick fit of the data from Raju *et al.* [30] to both linear and $T^2$ trends yield similar fitting statistics which heavily depend on the region of fit (Figure 12d). Alternatively, differences in the heat capacity between single crystal and powder samples occur quite frequently and are normally caused by differences in sample quality, stoichiometry, or crystallinity (between different polycrystalline samples). However, we stress that this is not the underlying cause in the present case as all other measurements done on these crystals, especially the magnetic susceptibility, which is particularly affected by defects in Mo pyrochlores [9, 45], are consistent with all other studies done to date. Disorder at the bulk level is therefore an unlikely culprit leaving surface effects, crystallinity and domain size as more likely options. A small subsample of crushed single crystal was used for heat capacity measurements (Figure 12a). Not only does the $T^2$ behavior remain unchanged, but the low-temperature heat capacity remains unaltered upon annealing the crushed sample in $O_2$ at 300°C for 24 hours, which is below the temperature at which $MoO_3$ volatilizes from the surface. The only visible difference is at higher temperatures where mass errors play a larger role in the relative scatter. This strongly suggests that domain size, crystallinity, and surface effects have a negligible effect on the heat capacity of this system at low temperatures and provide evidence that the discrepancy is, in fact, caused by differences in instrumentation. It is therefore believed that the $T^2$ behavior of the heat capacity observed here is the true behavior of this system in both powders and single crystals, providing yet another new and crucial piece of evidence that was missed in earlier studies.

$T^2$ temperature dependent heat capacities have been observed before and are predicted for samples that have two-dimensional character with a linear dispersion of excitations [103], linear nodes on the Fermi surface [104], or orbital glass states [105] although the latter is quite rare. In particular, a comparison with $FeCr_2S_4$ warrants some discussion. $FeCr_2S_4$ adopts the spinel structure ($Cr^{3+}$ ions occupy the same frustrated sublattice as $Mo^{4+}$ in $Y_2Mo_2O_7$). Single crystals of $FeCr_2S_4$ show a $T^2$ dependence in the low temperature lattice-subtracted heat capacity, in stark contrast to a lambda anomaly attributed to orbital ordering in powders [106, 107]. Furthermore, it was shown that orbital ordering in powders can be suppressed either by doping ions onto the B site [108] or simply through alternative methods of sample preparation [109]. Both the heat capacity of powder and single crystalline $FeCr_2S_4$ are nearly magnetic-field independent, which is not expected for changes in the spin system [106, 107]. Both samples show an enhanced linear term in the heat capacity at temperatures greater than the orbital ordering transition that is attributed to an orbital liquid state at higher temperatures [106, 107].

Despite sharing a $T^2$ dependence of the low-temperature lattice-subtracted heat capacity observed in both single crystalline $FeCr_2S_4$ and single crystalline $Y_2Mo_2O_7$, these two materials are actually quite dissimilar. For example unlike $FeCr_2S_4$, no evidence was found for an enhanced linear term in the heat capacity above $T_f$ in $Y_2Mo_2O_7$. Other than the discrepancy in the temperature dependence of the heat capacity below 10 K, the heat capacity curve in powders and single crystals is remarkably consistent (to our knowledge, the magnetic field dependence of this peak was never published for powders). The similarity of the $Y_2Mo_2O_7$ heat capacity to that of $FeCr_2S_4$ orbital glass provides strong motivation to continue investigating the effect of spin-orbital coupling in $Y_2Mo_2O_7$, which has seldom been discussed in the literature albeit with one notable exception [41]. Naively, if spin-orbital coupling truly plays a role in the spin freezing, one might even expect a $T^2$ dependence in the heat capacity rather than a linear dependence from extra degrees of freedom. However, we should not be too quick to draw any meaningful conclusions solely from the heat capacity. The exchange interactions in sulfides are inherently different than oxides. Additionally, the $FeCr_2S_4$ spinel contains two 3*d* magnetic ions while the pyrochlore $Y_2Mo_2O_7$ contains one 4*d* magnetic ion. Whereas magnon contributions are negligible for the low-temperature heat capacity of $FeCr_2S_4$, a small, but observable spin component does contribute to the low-temperature heat capacity of $Y_2Mo_2O_7$ that cannot be separated out using a simple lattice subtraction.

**E. Density functional theory calculations**

The importance of orbital degrees of freedom in $Y_2Mo_2O_7$ and their coupling to spin and lattice degrees of freedom can be theoretically demonstrated using *ab initio* electronic structure calculations. We investigated the electronic behavior of $Y_2Mo_2O_7$ for two different configurations of Mo spins: ferromagnetic and antiferromagnetic (Mo atoms 1 and 4 with spin up and Mo atoms 2 and 3 with spin down, Figure 13, see Figure 14 for labeling legend). While neither of these spin-ordered states is the true ground state of $Y_2Mo_2O_7$, important implications can be drawn regarding microscopic processes that may be occurring in $Y_2Mo_2O_7$. Different orbital ordering for the two magnetic configurations is observed indicating a strong coupling between the Mo orbital and spin degrees of freedom. This is a natural consequence of a combined effect of crystal field splitting and the electronic configuration of the $Mo^{4+}$ ion. Indeed, one of the two Mo electrons occupies the lowest $a_{1g}$ state while the second electron is shared between the higher-lying degenerate $e'_g$ states [41, 110]. Under small perturbations like spin superexchange energy, on-site Coulomb repulsion, spin-orbit coupling or lattice distortions, various combinations of the $e'_g$ states can be occupied (the orbital states shown in Figure 13 being two such states). Moreover, each of these orbital ordered states is accompanied by local lattice distortions, which, as our *ab initio* structural relaxations reveal, are mostly variations in the Mo-O and Y-O distances (Table 1). Physical evidence of such variations in the local structure is suggested by previous NMR [35, 36], μSR [29], and nPDF [38] experiments. It is also conceivable that distortions in the local structure could be responsible for the Huang scattering observed here, yet this remains purely speculative for now.

**IV. CONCLUSIONS**

Why does $Y_2Mo_2O_7$ show spin glass behavior? Previous theoretical studies have attributed the freezing to anisotropy due to orbital ordering and local distortions although both mechanisms are not necessarily mutually exclusive [56]. However, the degeneracy calculated by us and observed in this system is quite remarkable in that it may also extend from the spin to the orbital regime. Such an orbitally degenerate state has never been predicted for $Y_2Mo_2O_7$. The advantage of $Y_2Mo_2O_7$ over systems such as $FeCr_2S_4$ is that well-ordered crystals large enough for probes like neutron scattering exist. It would be instructive to perform polarized neutron scattering on the inelastic excitation spectrum in order to directly and unambiguously observe the effects of spin-orbital coupling. Furthermore, angle resolved photo-emission spectroscopy would allow for the complete characterization of the phonon spectrum so that the heat capacity can be adequately modeled. For an orbital glass, one would expect ill-defined low-energy excitations as a result of the degeneracy in the orbital ordering. As shown in Figures 6c (and the supplementary information), a broad isotropic excitation exists above the freezing temperature at energies below 2 meV, in line with the broad peak observed in the heat capacity. These excitations exist well above the Curie-Weiss temperature and may be indicative of the low-energy vibrations one would expect for an orbital glass, although a more comprehensive study of the excitations is required in the future. Thermal expansion measurements are also warranted; if local lattice distortions are truly responsible for the spin freezing, an anomaly might occur at the glassy transition temperature using such a precise technique [111] on a high quality single crystal that might otherwise be missed with diffraction. Looking back at the analogy with $FeCr_2S_4$, we can compare it to a similar compound, $FeSc_2S_4$ ($\theta$=-45 K), where neither spin nor orbital ordering occurs down to at least 50 mK [112]. The contrasting behavior between these two compounds was attributed to the unique hybridization of the Fe *d*, Cr/Sc *d*, and S *p* orbitals. One would expect stronger spin-orbital-lattice coupling in a Mo 4*d* compound, although we note that from our non-relativistic DFT calculations, the major contribution to the spin-orbital coupling is of the Kugel-Khomski type. Orbital interactions are anisotropic, frustrated, and are strongly coupled with lattice distortions. If our system *also* has strong antiferromagnetic spin interactions coupled, as we have shown, to both orbital and lattice degrees of freedom, then we can naively speculate that the glassy behavior is a consequence of the failure of the system to satisfy the strongly interacting frustrated spins, the frustrated orbital interactions *and* orbital degeneracy simultaneously.

In summary, we have grown single crystalline $Y_2Mo_2O_7$ and have characterized it with magnetization, heat capacity, neutron, and X-ray scattering techniques. The observed rings of scattering strongly suggest spin liquid-like correlations within this unique spin glass despite having a well-ordered crystalline structure. We have shown using *ab initio* density functional theory calculations that orbital degrees of freedom are an important ingredient to the physics at play in this system – the new $T^2$ heat capacity dependence also hints at a degenerate orbital component that has been ignored in previous discussions.

**Acknowledgements**

This work has been supported by NSERC, the ACS Petroleum Fund, the CRC program, CFI and the DFG (grant SFB/TRR49). H.J.S gratefully acknowledges support from the Vanier CGS (NSERC) and MGS programs as well as the University of Manitoba. In addition to NSERC, M.J.P.G. would like to thank support from the CRC Program. J.A.M.P. gratefully acknowledges funding from the STFC and EPSRC (EP/G004528/2). The authors want to acknowledge useful discussions with A. B. Dabkowski, K. McEleney, Z. Islam, Y. Feng, M. Bieringer, J. Van Lerop, H. Takagi, and J.E. Greedan. The NHMFL is operated under a cooperative agreement with Florida State University and the NSF under DMR-0654118. This work utilized facilities supported in part by the NSF under Agreement No. DMR-0944772. A portion of this research at ORNL's SNS and Argonne National Laboratory's APS was sponsored by the Scientific User Facilities Division, Office of Basic Energy Sciences, U.S. Department of Energy (APS under Contract No. DE-AC02-06CH11357). We are greatly appreciative of the staff and support of the NRC at Chalk River Laboratories.

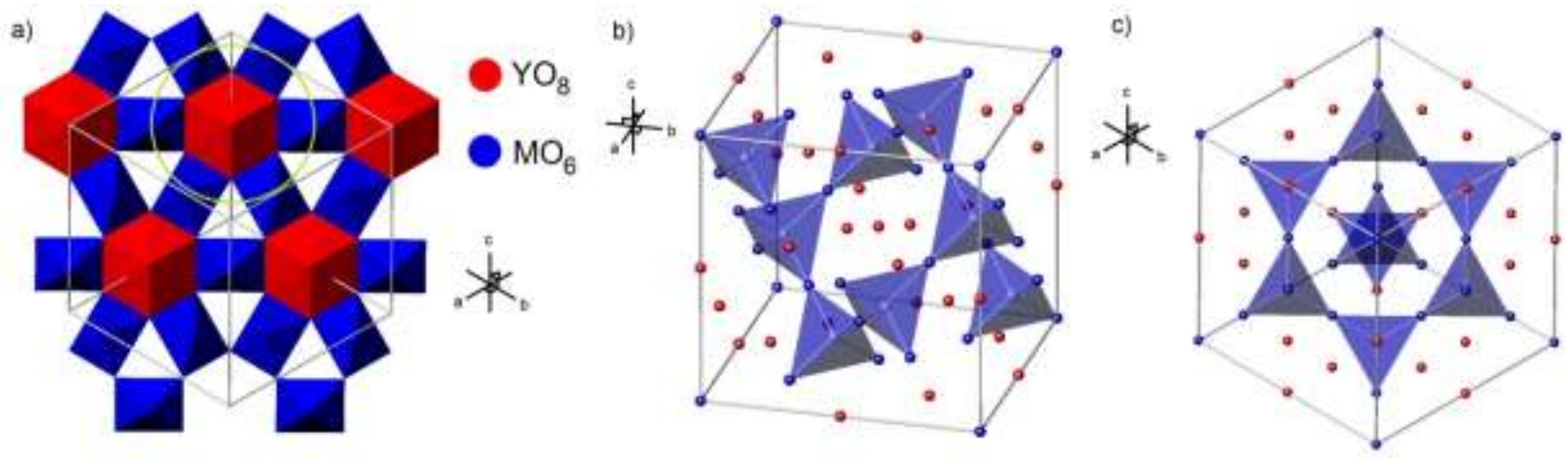

**Figure 1:** The crystal structure of $Y_2Mo_2O_7$ (S.G. Fd-3m) can be thought of in a number of different ways: **a)** A single layer of 8-fold coordinated Y-O and 6-fold coordinated Mo-O polyhedra share edges such that one can build the structure through ABCABC close-packing of the area encircled in green along <111>; **b)** Two interpenetrating corner sharing tetrahedral networks with cationic vertices and vacancy centers. In this figure, only the Mo-tetrahedral network is shown (the Y-network is displaced by k=(0.5, 0.5, 0.5)); **c)** The tetrahedral networks form a pseudo-close-packing network of kagomé planes along <111>.

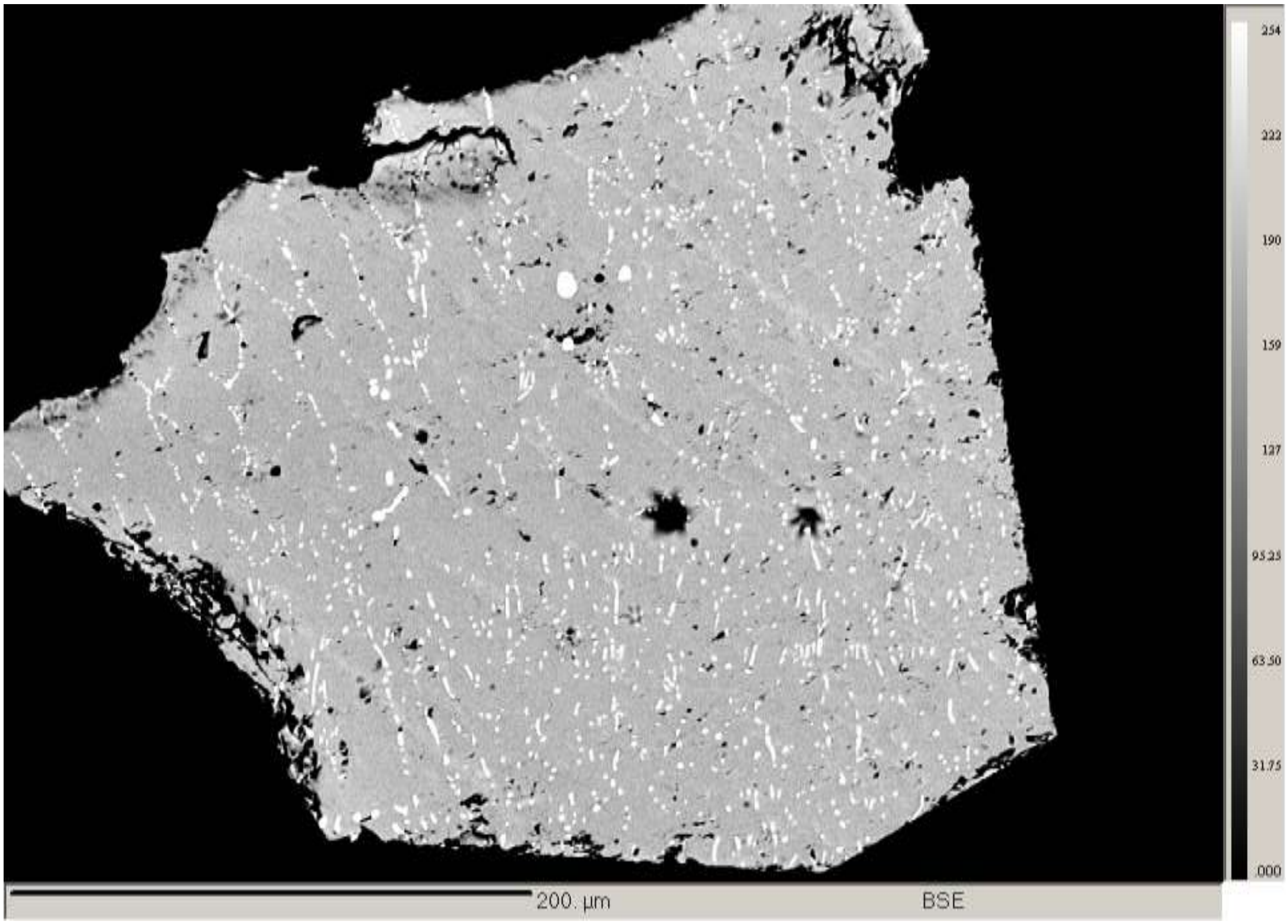

**Figure 2**: Back-scattered electron image a small cross-section of single-crystal $Y_2Mo_2O_7$ produced by wavelength-dispersive electron microprobe analysis. The grey area is stoichiometric $Y_2Mo_2O_7$ while the white areas are small Mo metal inclusions as a result of a sputtering process used to probe the interior. These metal inclusions are a surface impurity only detected on this particular subsection of the crystal. Thus, they are not detected in any of our magnetic susceptibility, heat capacity, synchrotron X-ray, or neutron diffraction measurements.

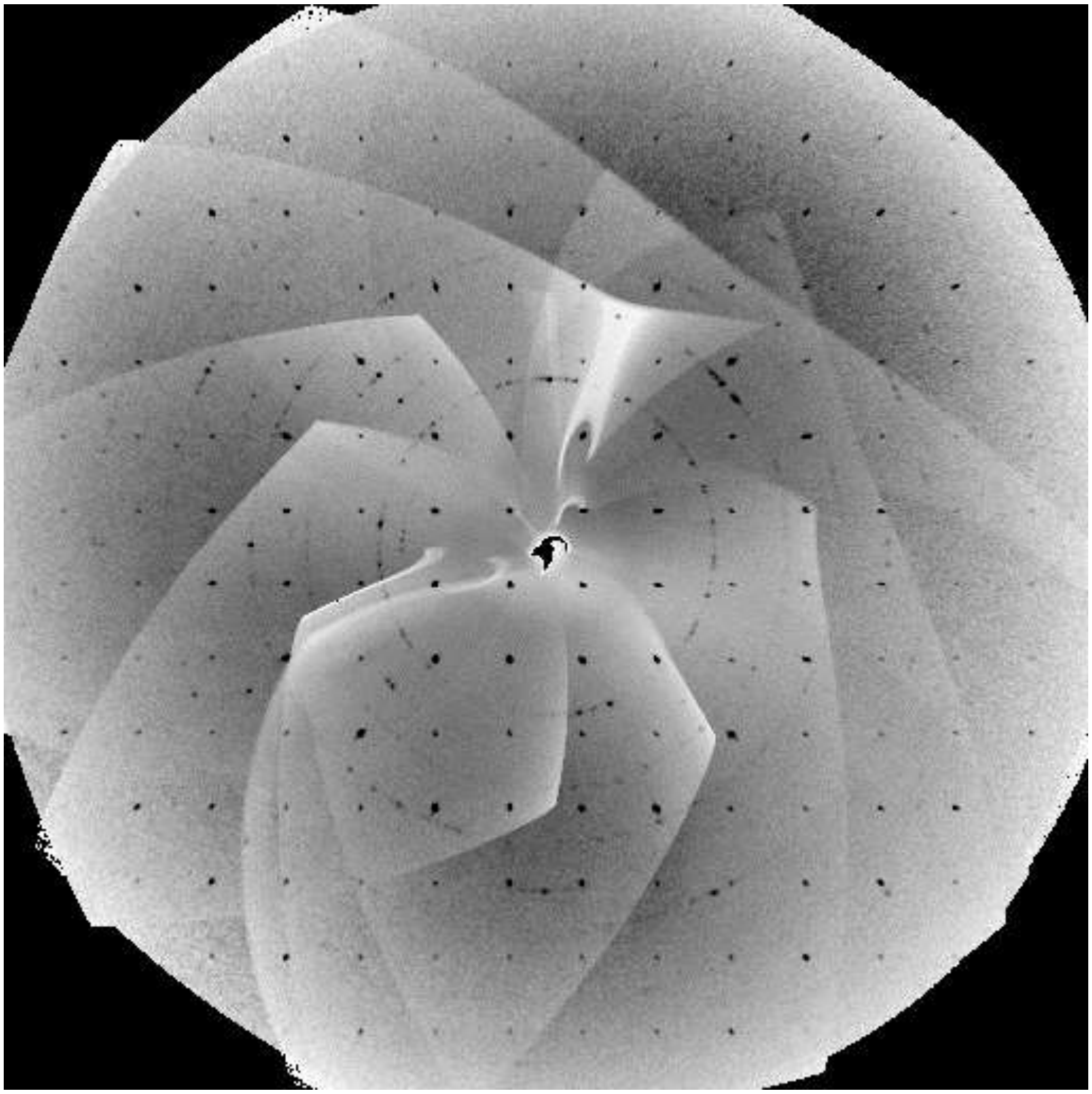

**Figure 3:** Composite image of X-ray diffraction exposures taken using CCD imaging. The black spots are from the dominant $Y_2Mo_2O_7$ phase refined to ICSD: 202522. It was determined that our 0.03 g single crystal subsection contained a small collection of grains about 0.1 mm in size. The lattice constant was determined to be 10.28 Å, which is slightly higher than reported values of 10.230(1) Å [26]. While a larger lattice constant can be due to O deficiencies within the crystal, our annealing step in the preparation, electron microprobe, and magnetic susceptibility suggest that this is not the case. The slightly larger lattice constant is instead due to the composite nature of crystal at X-ray penetration depths.

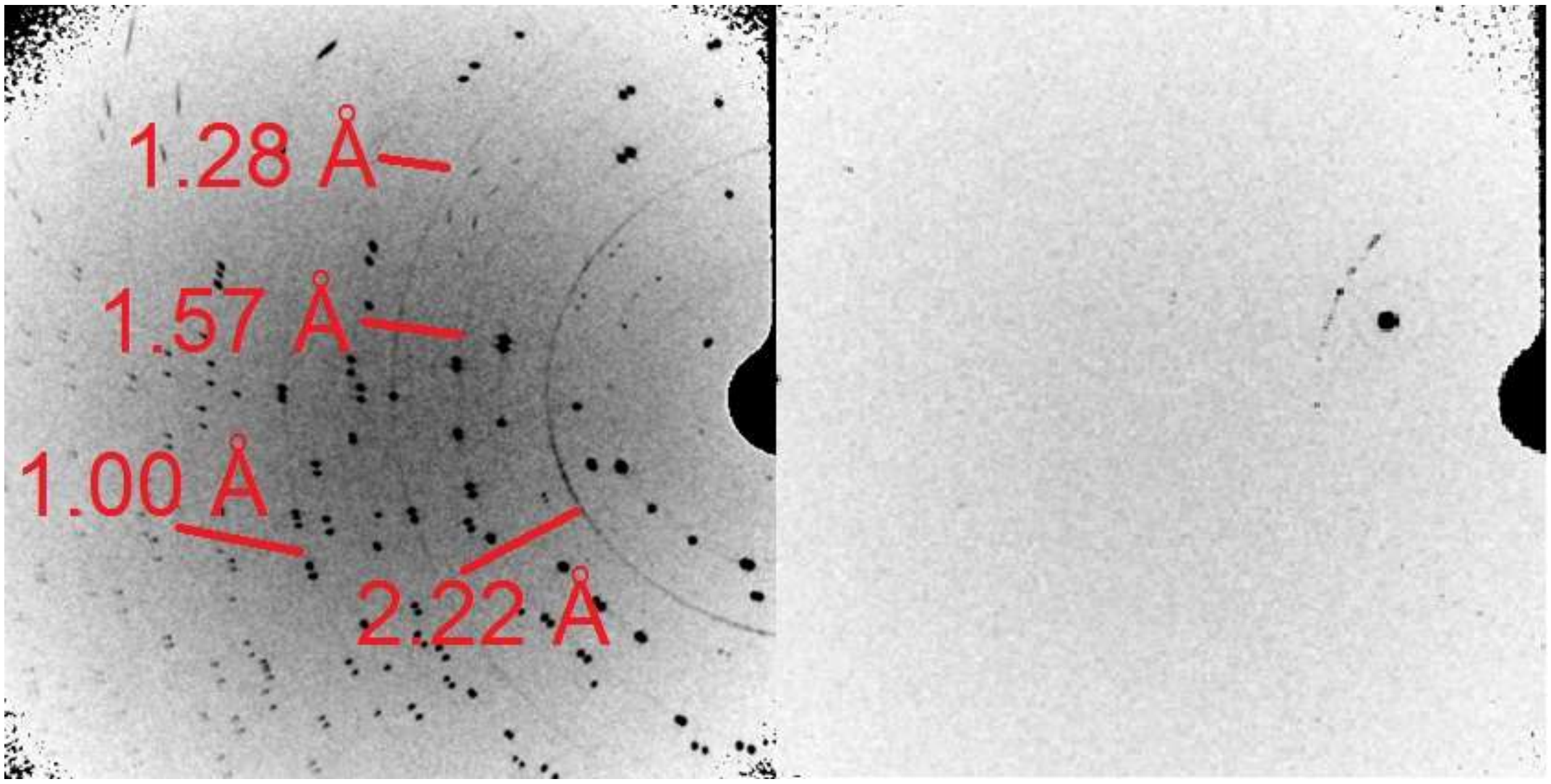

**Figure 4:** Individual CCD exposures at 30 s **(left)** and 3 s **(right)**. The Mo metal arcs have been labeled in red.

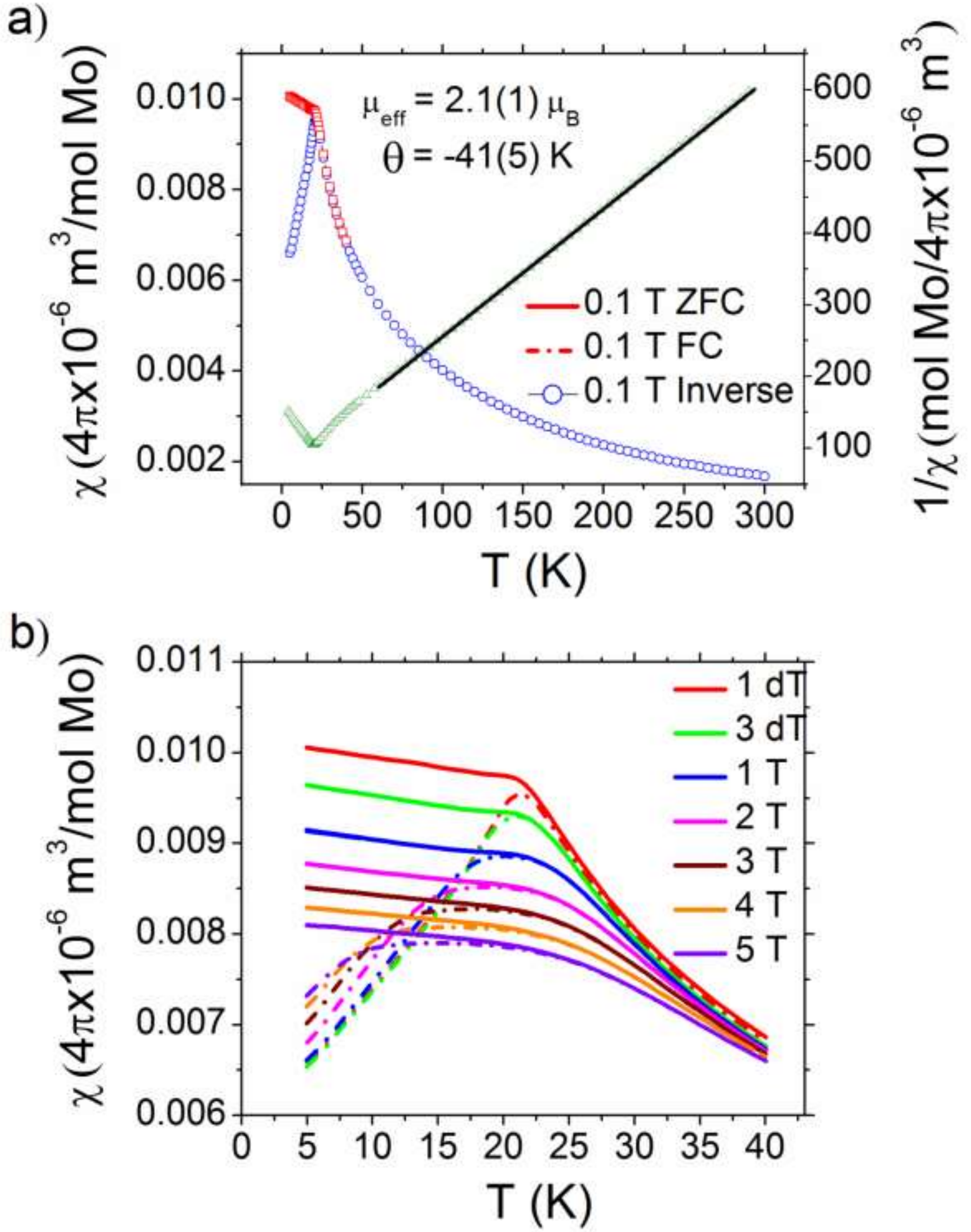

**Figure 5:** For all relevant panels, error bars are smaller than the symbol size; **a)** DC-susceptibility and inverse susceptibility of $Y_2Mo_2O_7$ fit to the Curie-Weiss Law using a field of 1 T applied along [111]; **b)** The field-dependence of $T_f$ as observed using DC-susceptibility.

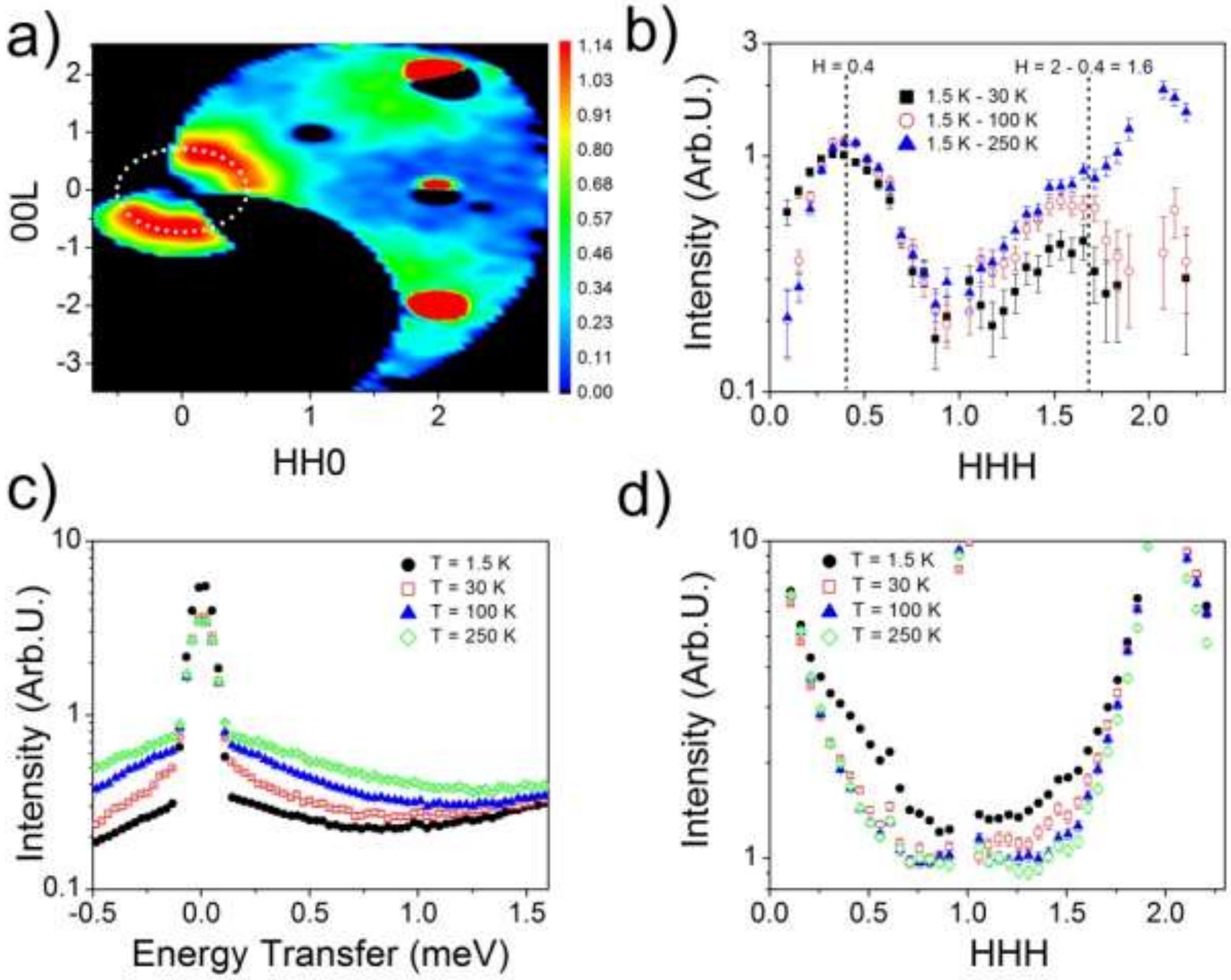

**Figure 6: a)** The elastic ring observed using neutron scattering. Data from 100 K was subtracted from data at 1.5 K and binned over energy E = [-0.12,0.12] meV. The data has been smoothed in this figure to better show the broad features of the ring. The white points are calculated for constant Q = 0.44 Å$^{-1}$; **b)** The ring is replicated at the center of the next Brillouin Zone. Data has been binned over L = [-0.1,0.1] r.l.u.; **c)** Inelastic data integrated over HH0 = [0,1] r.l.u.and 00L = [-1,0] r.l.u. (error bars are smaller than the symbol size); **d)** Raw data depicting the evolution of the rings. Peaks at 0.6 and 1.4 r.l.u. are spurious features of the instrument. Note that the data is shown on a logarithmic scale due to the enormous intensity difference between the (222) Bragg peak and the diffuse scattering.

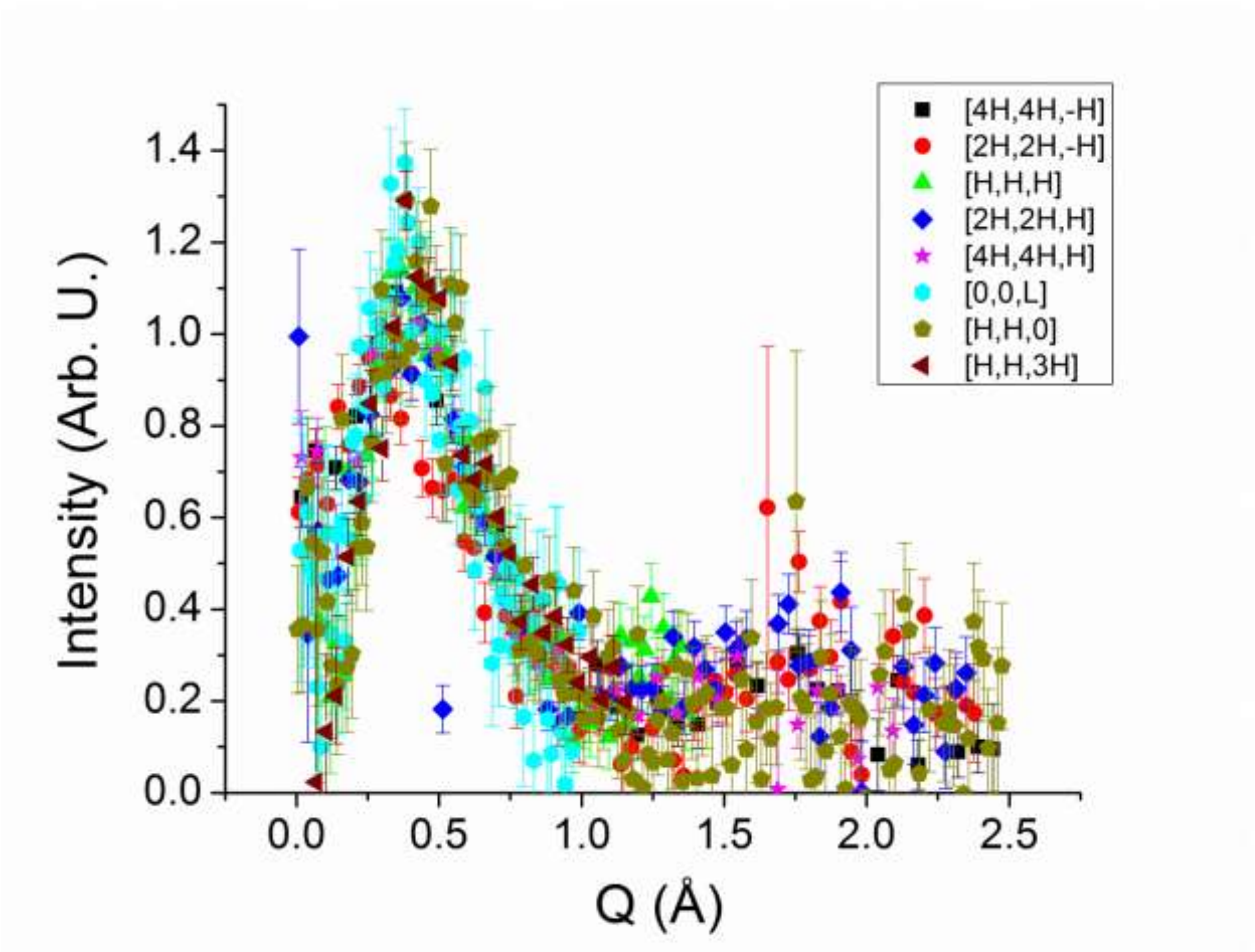

**Figure 7:** Cuts along various directions through the ring. Data taken at 100 K was subtracted from data at 1.5 K. All cuts were made integrated over the elastic peak with step size 0.02 r.l.u. The data along [2H,2H,-H] appears to break trend due to a lack of detector coverage along that cut.

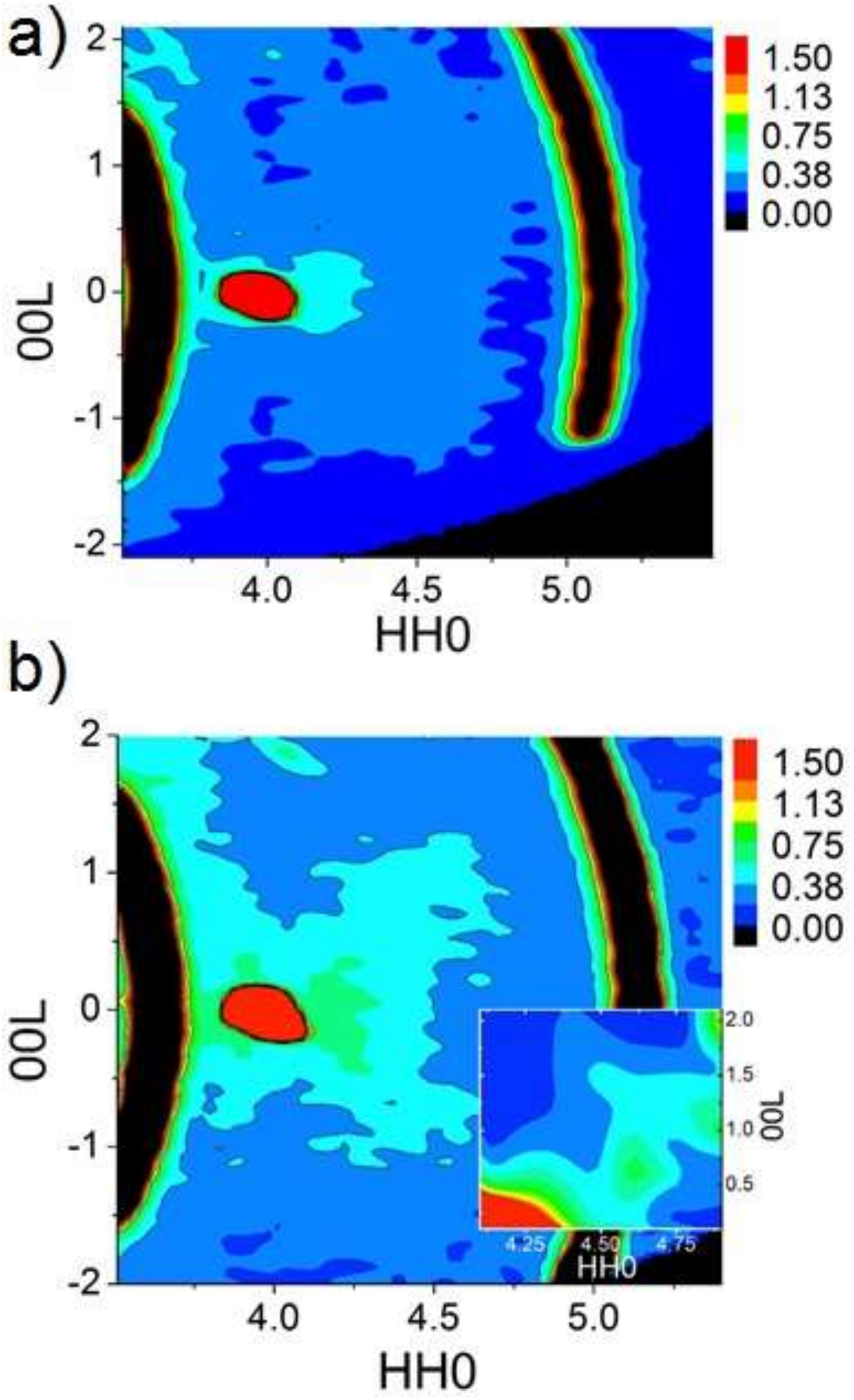

**Figure 8:** Single crystal neutron diffraction ($E_i$ = 20 meV) integrated over E = [-0.02, 0.02] meV and H-H0 = [-0.2, 0.2] r.l.u. taken on the CNCS at 300 K **(a)** and 1.5 K **(b)**. The Huang scattering and temperature dependence (not shown) is also observed using the C5 instrument (CNBC, Chalk River, **inset**) at 4 K. Unlike the pattern observed with X-rays, temperature dependence is observed, which might indicate spin-orbital coupling. Aluminum powder lines have been artificially colored black for clarity.

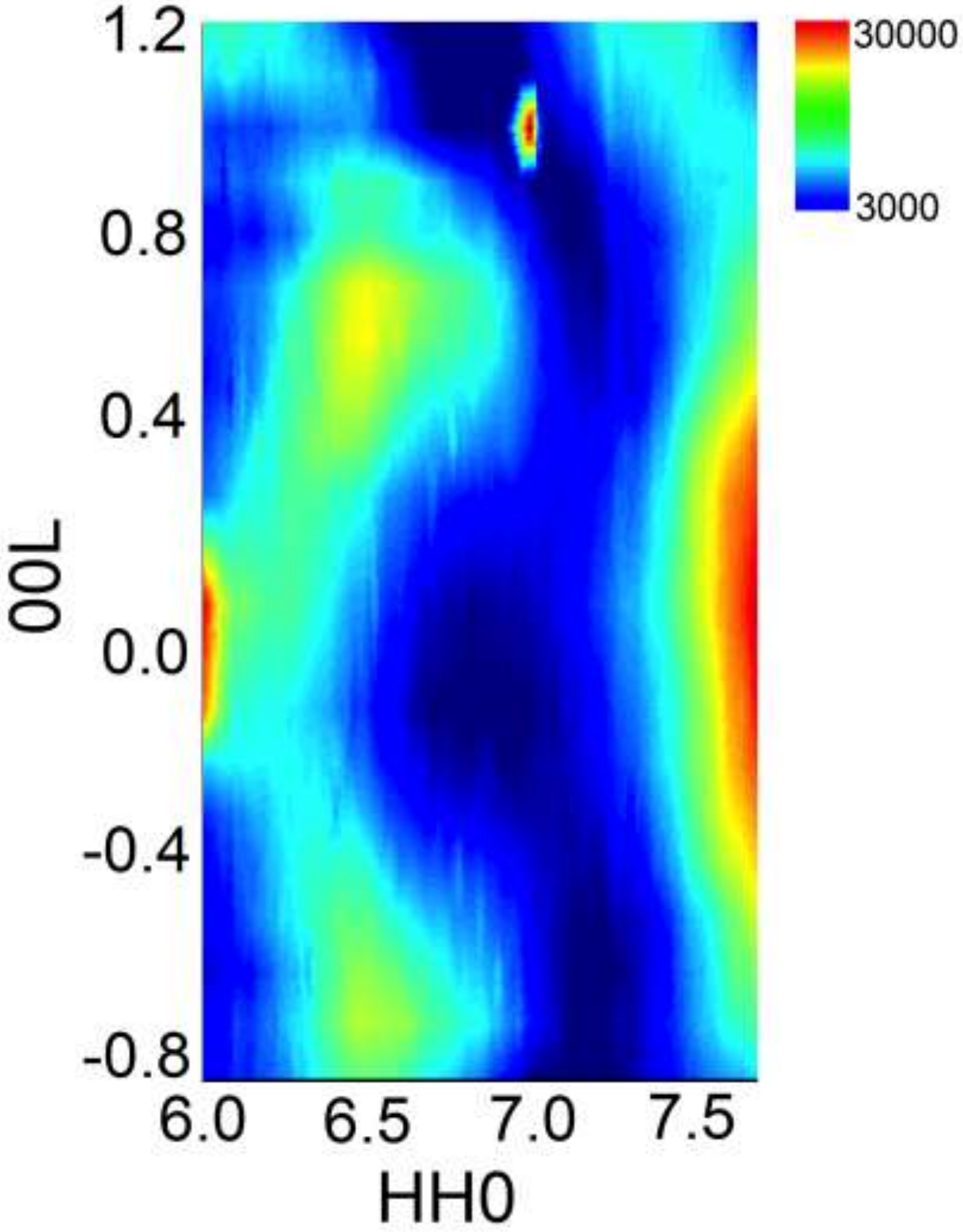

**Figure 9:** Example of Huang scattering as observed with synchrotron X-rays at 10 K. The intensity is on a logarithmic scale. In addition to the observed feature at (660), Huang scattering orders of magnitude more intense was also observed on the (440), (666), (008) and (880) peaks, the latter of which is visible on the right here. But in general, this Huang scattering is extremely weak and was not observed using a two-circle laboratory source equipped with 14.4 keV X-rays. No obvious temperature dependence was observed; the scattering persisted at all temperatures between 10 and 300 K.

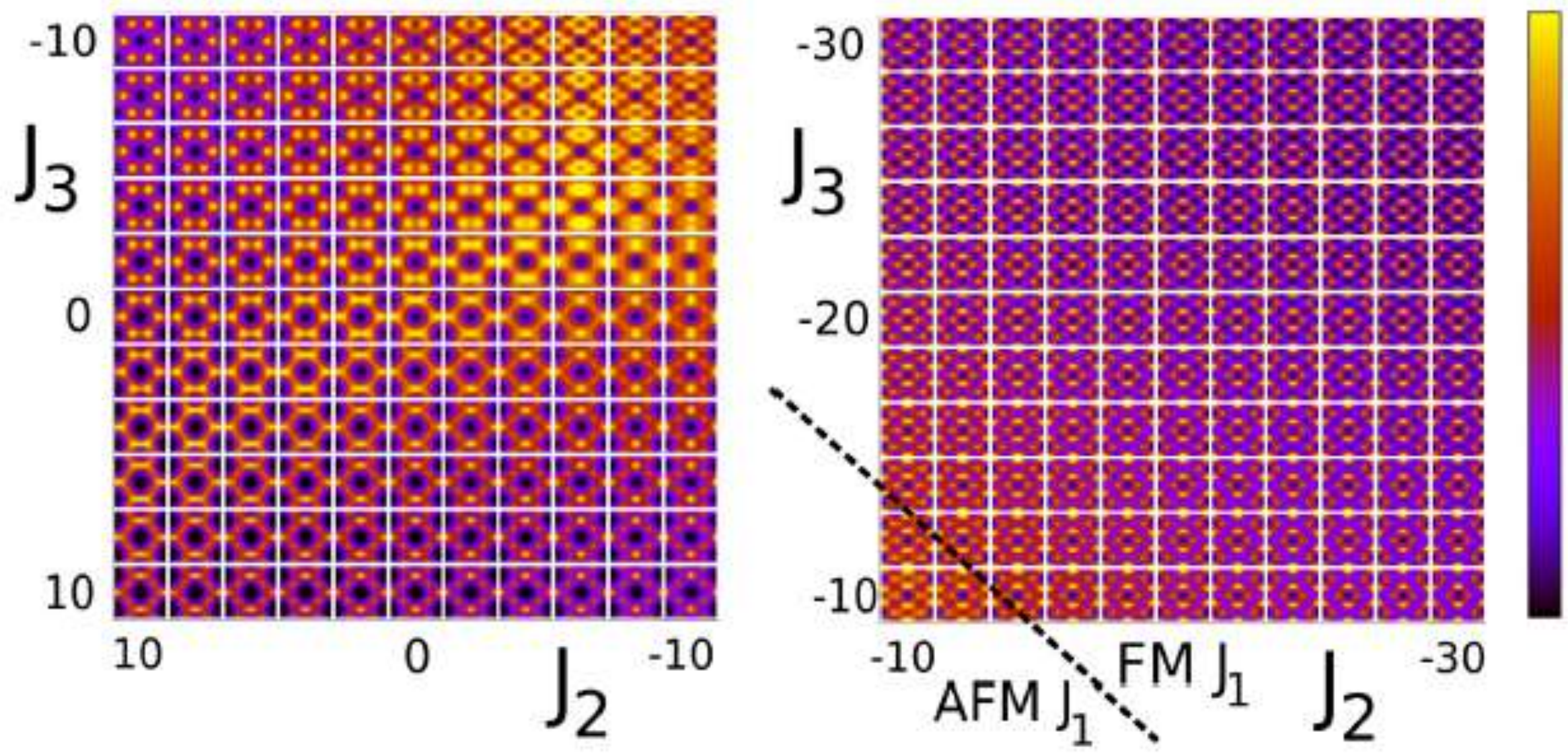

**Figure 10:** Neutron scattering plots in the HHL plane for $Y_2Mo_2O_7$ at T = 150 K, with a range of next nearest neighbor terms $J_2$, $J_3$ (in Kelvin). Each plot covers H ∈ [-2, 2], L ∈ [-3, 3]; the arbitrary scale of each plot is normalized to the color bar on the right. The Curie-Weiss constraint $4(J_1 + 2J_2 + 2J_3)$ = -200 K is applied, meaning $J_1$ is ferromagnetic for $J_2 + J_3 <$ -25 K (marked in the right hand diagram).

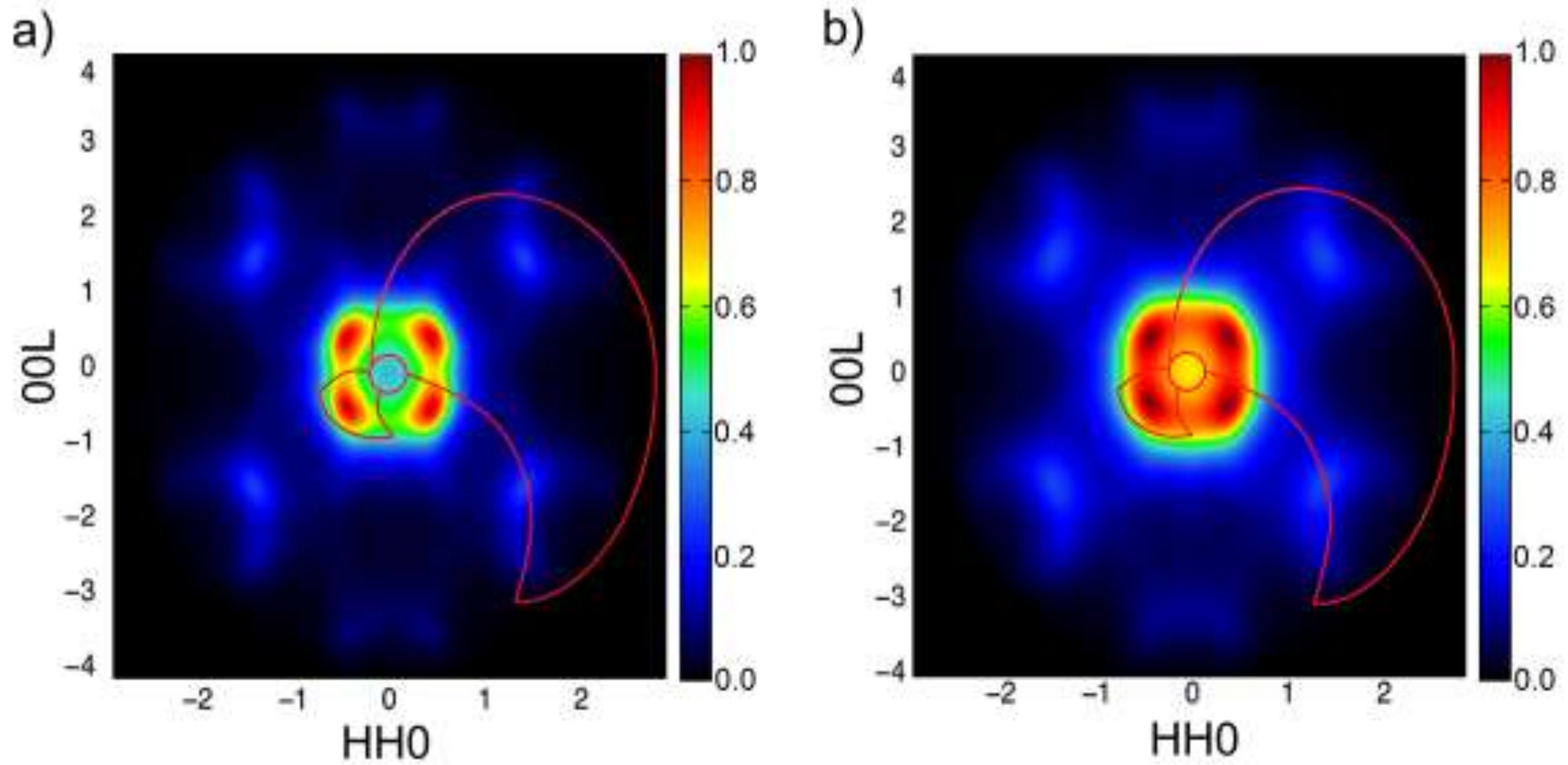

**Figure 11: a)** Large N Heisenberg model of the diffuse scattering $J_{nnn}/J_{nn}$=-0.25 at T=0.6*$J_{nn}$ ; **b)** Mean field theory Heisenberg model at T=1.2$T_c$ (same couplings) where $T_c$ is the mean-field model transition temperature. Note that this choice of couplings fails to fulfill $\theta_{CW}$<0 K. The red and black outlines in each figure represent the data coverage of the spectrometer used in Figure 6.

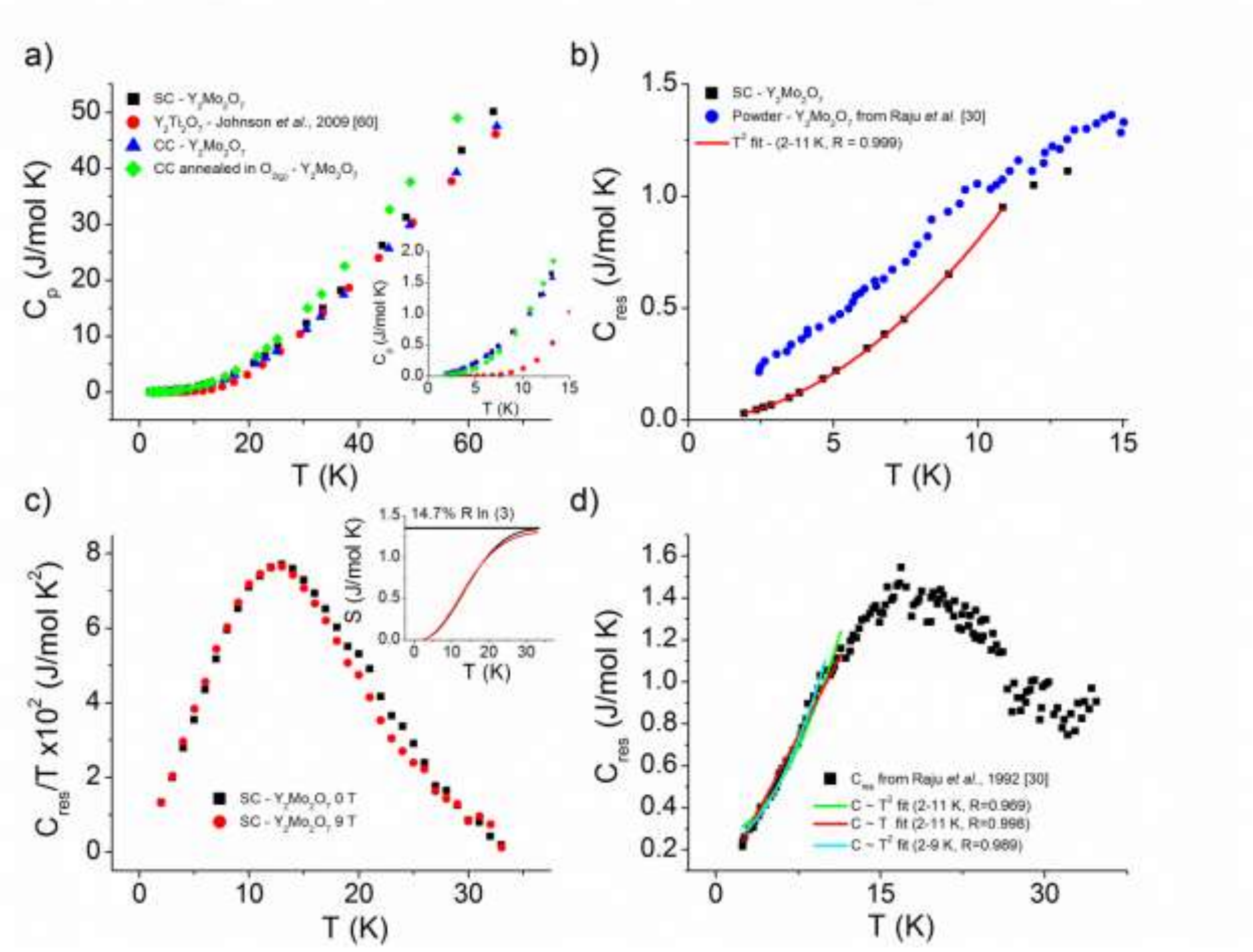

**Figure 12:** Error bars for all measurements on single or crushed crystals of $Y_2Mo_2O_7$ are smaller than the symbols. **a)** Raw heat capacity data of single crystal $Y_2Mo_2O_7$ (black square), crushed crystals of $Y_2Mo_2O_7$ (blue triangle), crushed crystals annealed in $O_{2(g)}$ for 24 hours at 300°C (green diamond), and $Y_2Ti_2O_7$ from Johnson *et al.*, 2009 [60] (red circle). Differences between $Y_2Mo_2O_7$ samples at higher temperatures are due to a mass errors (or surface effects for the annealed crushed crystals) while no clear difference is discernible at lower temperatures **(inset)**; **b)** The low temperature heat capacity of single crystal $Y_2Mo_2O_7$ is fit to a strict $T^2$ law (red curve) and is compared to powder samples (Raju *et al.*, 1992 [30]); **c)** Field dependence of single crystal $Y_2Mo_2O_7$ lattice-subtracted heat capacity data using our own sample of $Y_2Ti_2O_7$ (found to be consistent with previous experiments [60]). The field was applied along the [111] direction. Integrating the peak should yield the entropy released by this system at low temperatures **(inset)**. Only 14.7% of the theoretical maximum entropy (9.13 J/mol K) is released; **d)** Heat capacity data from Raju *et al.* [30] is quite noisy probably due to instrumental effects in the subtraction. The low temperature data can be fit to both a T (red curve) and $T^2$ dependence (blue and green curves) within a reasonable margin of error dependent on the region of fit.

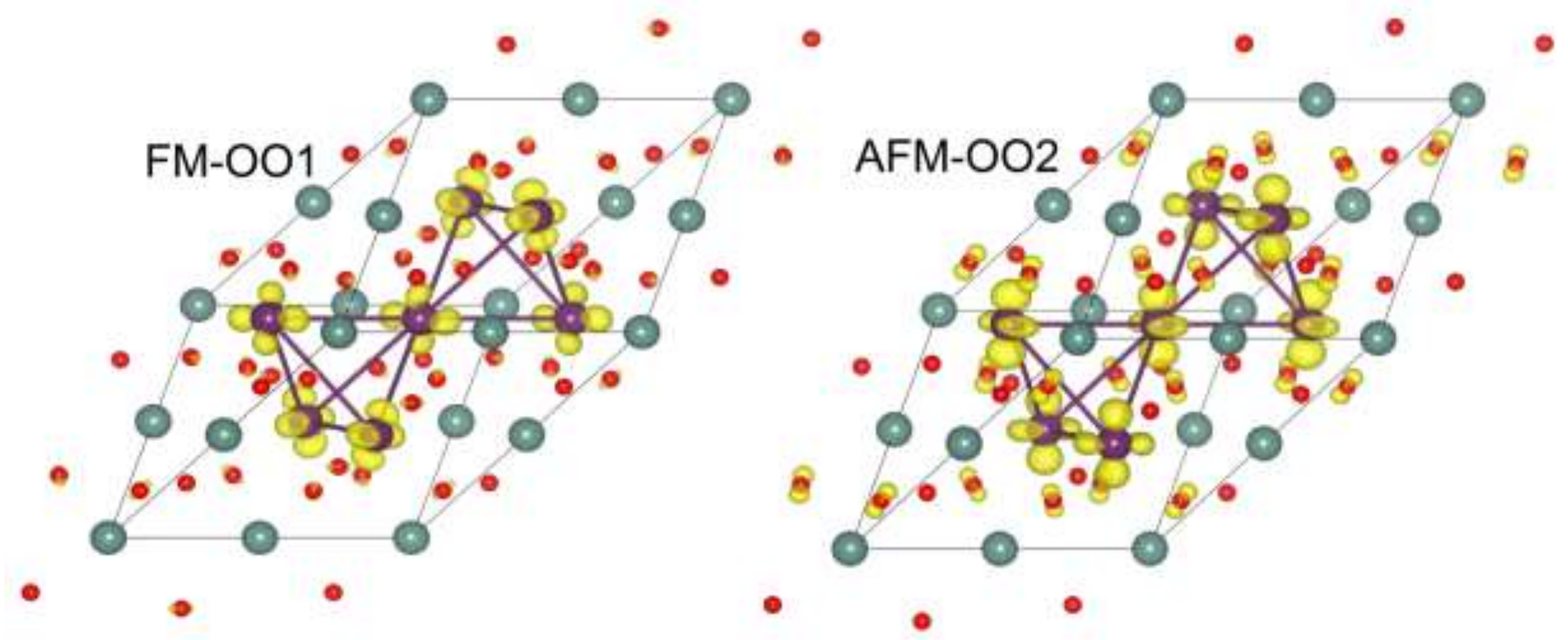

**Figure 13:** $Y_2Mo_2O_7$ unit cell with four Mo atoms (space group *P*-1). Density iso-surfaces of the valence electrons are shown in yellow; one clearly distinguishes the different occupations of Mo valence orbitals in the ferromagnetic spin order-orbital order 1 (FM-OO1) and antiferromagnetic spin order-orbital order 2 (AFM-OO2) solutions.

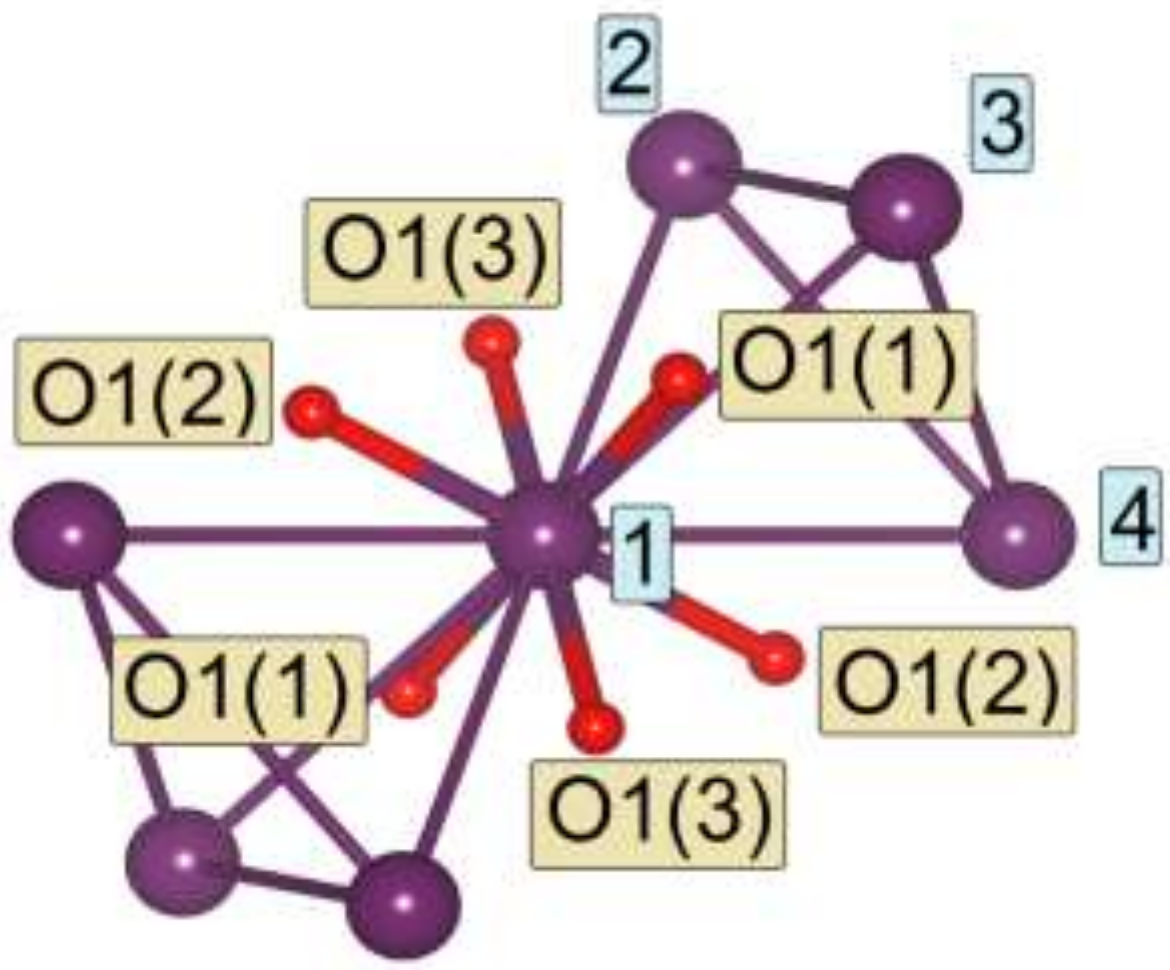

**Figure 14:** Legend for Table 1. Mo atoms are labeled by the blue shaded boxes.

**Table 1:** Interatomic distances in the experimentally determined $Y_2Mo_2O_7$ structure [26] and in the density functional theory relaxed structures with the FM-OO1 and AFM-OO2 configurations. Oxygens O1(1), O1(2) and O1(3) are indicated in Figure 14.

| **Bond** | **Distance (Å)** | | |
|---|---|---|---|
| | *Ref. [26]* | *FM-OO1* | *AFM-OO2* |
| Y-O1(1) | 2.452 | 2.383 | 2.419 |
| Y-O1(2) | 2.452 | 2.442 | 2.454 |
| Y-O1(3) | 2.452 | 2.489 | 2.419 |
| Y-O2 | 2.215 | 2.214 | 2.215 |
| Mo-O1(1) | 2.021 | 2.006 | 2.044 |
| Mo-O1(2) | 2.021 | 2.031 | 2.019 |
| Mo-O1(3) | 2.021 | 2.058 | 2.042 |
| Mo-Mo | 3.617 | 3.617 | 3.617 |
| Y-Y | 3.617 | 3.617 | 3.617 |